\RequirePackage{fix-cm}
\documentclass[twocolumn]{svjour3}   
\smartqed 
\usepackage{cite}
\usepackage{balance}
\usepackage{graphicx}
\usepackage[cmex10]{amsmath}
\usepackage{xfrac}
\usepackage{algorithmic}
\usepackage{array}
\usepackage[toc,page]{appendix}
\usepackage{color, colortbl}
\definecolor{Gray}{gray}{0.9}
\definecolor{LightCyan}{rgb}{0.8,1,1}
\newcolumntype{g}{>{\columncolor{Gray}}l}
\usepackage{multirow}
\usepackage{subfigure}
\usepackage{url}
\usepackage{balance}
\usepackage{color, colortbl}
\definecolor{Gray}{gray}{0.9}
\definecolor{LightCyan}{rgb}{0.8,1,1}
\newcolumntype{g}{>{\columncolor{Gray}}l}
\usepackage{array}
\newcolumntype{L}[1]{>{\raggedright\let\newline\\\arraybackslash\hspace{0pt}}m{#1}}
\newcolumntype{C}[1]{>{\centering\let\newline\\\arraybackslash\hspace{0pt}}m{#1}}
\newcolumntype{R}[1]{>{\raggedleft\let\newline\\\arraybackslash\hspace{0pt}}m{#1}}
\begin{document}
\title{Design of Configurable Sequential Circuits in Quantum-dot Cellular Automata} 
\titlerunning{Design of Configurable Sequential Circuits in Quantum-dot Cellular Automata}      
\author{Mrinal Goswami \and Mayukh Roy Chowdhury \and Bibhash Sen}
\authorrunning{M. Goswami et. al.} 
\institute{M. Goswami, M. R. Chowdhury and B. Sen \at
              Department of Computer Science and Engineering, \\National Institute of Technology, Durgapur,\\ West Bengal - 713209, India \\
              \email{mg.16cs1102@phd.nitdgp.ac.in}}           
\date{Received: date / Accepted: date}
\maketitle
\begin{abstract}
Quantum-dot cellular automata (QCA) is a likely candidate for future low power nano-scale electronic devices. Sequential circuits in QCA attract more attention due to its numerous application in digital industry. On the other hand, configurable devices provide low device cost and efficient utilization of device area. Since the fundamental building block of any sequential logic circuit is flip flop, hence constructing configurable, multi-purpose QCA flip-flops are one of the prime importance of current research. This work proposes a design of configurable flip-flop (CFF) which is the first of its kind in QCA domain. The proposed flip-flop can be configured to D, T and JK flip-flop by configuring its control inputs. In addition, to make more efficient configurable flip-flop, a clock pulse generator (CPG) is designed which can trigger all types of edges (falling, rising and dual) of a clock. The same CFF design is used to realize an edge configurable (dual/rising/falling) flip-flop with the help of CPG. The biggest advantage of using edge configurable (dual/rising/falling) flip-flop is that it can be used in 9 different ways using the same single circuit. All the proposed designs are verified using QCADesigner simulator.
\end{abstract}
\section{Introduction}\label{sec:introduction}
CMOS based VLSI designs play a very important role in digital design. The operation of computer chips has shown outstanding growth in the last few decade. But the exponential increase of computing power over the last decades has relied on shrinking the transistor. The recent survey of International Technology Roadmap for Semiconductors (ITRS-2015) shows that the transistor could face the challenge of feature size limitation by 2021 \cite{itrs}. Quantum-dot Cellular Automata (QCA) is an emerging nanotechnology with extremely small feature size and ultra low power consumption which is suggested as an alternative technology of current CMOS \cite{lent01}.
\par Flip-flop plays a significant role in any sequential circuits. Sequential circuits in QCA gain maximum attention due to its realization difficulties in terms of architectural complexity and synchronous mechanism \cite{Abutaleb2017}. Till date, various studies have been performed on flip-flop design. However, all the previous designs realized different flip-flops with disparate designs and thus they differ in terms of area and latency. No regular, uniform or symmetric design paradigm is addressed so far for a simple implementation of flip-flops in QCA. The disparity of designs (SR, D, T and JK) leads to various problems in a single circuit having multiple flip-flops. The inter-connection of the different flip-flops (SR, D, T and JK) is another big issue. So, to tackle these issues, a uniform design methodology is to be explored for flip-flops having same area and latency.     
\par  In the recent past, conventional QCA based flip-flops were studied extensively \cite{Xiao2012,FF1,FF2,FF3,FF4,FF5,FF6} but, most of the previous designs are not robust and highly error prone due to defective wire-crossing employed in their designs \cite{defect}. Generally,  two well known wire-crossing techniques have been used in QCA technology: multi-layer and coplanar wire-crossing (using rotated QCA cells). Multi-layer wire-crossing has several fabrication limitations \cite{WC1} \cite{WC100} due to which it is not considered for our proposed designs. On the other hand, coplanar wire-crossing in QCA can be realized with $45^{0}$ (normal QCA cell) cells and $90^{0}$ (rotated QCA cell) cells. However, production of $90^{0}$ cell needs a high level of precision in placing which increases the overall cost and the implementation complexity \cite{WC1} \cite{WC100}.
\par  On the other hand, flexibility and performance are the two big issues of any digital logic circuit \cite{book1}. The intermediate trade-off between flexibility and performance is the utmost necessity for current digital logic circuit which can be attained by reconfigurability \cite{bondalapati01}. It is observed that reconfigurability in the nanoscale era can lead to the design of various low power and energy efficient system which is the need of the hour \cite{thesis01}.
\par All these above factors motivate us to design configurable, robust flip-flop structures for QCA which can address the irregular behaviour of designs as well as reliability issue of wire-crossing in QCA. The main contribution of this research is as follows:
\begin{itemize}
\item A configurable level triggered QCA flip-flop (\textbf{CFF}) is designed which can be configured to JK, D and T flip-flop. This is the very first attempt to design a configurable flip-flop in QCA.    
\item To remove the wire-crossing difficulties, clock-zone based coplanar crossover technique is employed, which is the most robust wire-crossing technique in QCA technology \cite{shin}\cite{abedi01}.
\item Considering CFF as a basic element, an edge configurable (dual/rising/falling) flip-flop is designed using a clock pulse generator (CPG).
\item Based on the edge configurable (dual/rising/falling) flip-flop, an n-bit edge configurable (dual/rising/ \\falling) counter/shift register is also proposed.
\end{itemize}
The rest of the paper is organized as follows: section \ref{Basic} introduces fundamentals of quantum-dot cellular automata (QCA). Existing works based on reconfigurable logic are discussed in section \ref{related}. The proposed design of level triggered configurable flip-flop (CFF) is introduced in section \ref{model}. The design of edge configurable flip-flop is introduced in section \ref{DETTT}. The higher order QCA configurable circuits are proposed in section \ref{counter/register} employing edge configurable flip-flop as the basic element. Finally, section \ref{Con}, concludes the paper.  
\section{Basics of QCA}\label{Basic}
\begin{figure}
\centering
\includegraphics[trim =0cm 0cm 0cm 0cm, clip,scale=0.30]{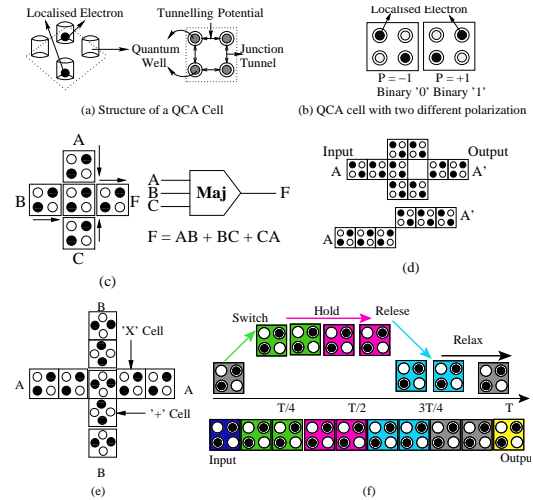}
\caption{The basic structures in QCA}
\label{basic}
\end{figure}
\begin{figure}
\centering
\includegraphics[trim =0cm 0cm 0cm 0cm, clip,scale=0.40]{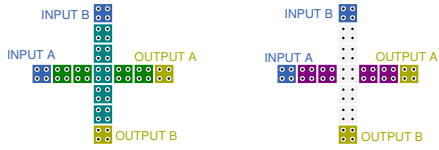}
\caption{The clock zone based wire-crossing \cite{shin}\cite{abedi01}}
\label{czb}
\end{figure}

A QCA cell comprises of four quantum dots (Figure \ref{basic}(a)) which can carry two free electrons. These free electrons can move between the four quantum dots. As shown in Figure \ref{basic}(b), the two polarization states of a QCA cell can be represented as P=-1 (logic 0) and P=+1 (logic 1). The basic structures of QCA technology are majority voter (Figure \ref{basic}(c)) and inverter (Figure \ref{basic}(d)). The QCA majority voter (MV) can be expressed as,
\begin{center}
MV(A, B, C) = AB + BC + CA
\end{center}
A majority voter can serve as a 2-input AND gate if one of the inputs is fixed at P=-1. Alternatively, if any one input of the majority voter is fixed at P=+1 then the modified majority voter can serve as a 2-input OR gate. There are two different types of inverter available in QCA technology as shown in Figure \ref{basic}(d). The possible two orientations ($``+"$) and ($``\times"$) ($90^ {0} $ and $45^ {0} $ respectively) of QCA cell are shown in Figure \ref{basic}(e). There are two fundamental wire-crossing techniques (co-planar and multi-layer) available in QCA. As shown in Figure \ref{basic}(e), the co-planar wire-crossing can be implemented with the mix combination of two different orientations ($90^{0}$ and $45^{0}$) of QCA cells. The multi-layer wire-crossing can be implemented with the help of two or more different layers. As discussed earlier, co-planar and multi-layer wire-crossing techniques are facing serious issues \cite{WC1} \cite{WC100} due to which we consider clock-zone based wire-crossing technique which can be implemented using non-adjacent clock zones (phase difference is equal to $180^{0}$)  on the same plane \cite{shin}\cite{abedi01} as shown in Figure \ref{czb}. This eliminates the problem of interference between the QCA cells.   
\par The QCA clock controls the flow of information within the circuit. The circuit information is carried from one end to another with the help of clock. The clock zones of QCA are distinct and $90^{0}$ phase shifted \cite{clock11} \cite{Karkaj2017}. At the time of computation, the previous clock zone must hold its output, which can be achieved by splitting the clock into four phases: switch, hold, release and relax \cite{clock10} which is shown in Figure \ref{basic}(f).
\par Four types of implementation technology exists in QCA: (a) metal-island \cite{metel01}; (b) semiconductor \cite{semi01}; (c) molecular \cite{mole01} \cite{mole02}; and (d) magnetic \cite{mag01} \cite{mag02}. The metal-island is the first implementation technology created to demonstrate the behaviour of QCA which requires cryogenic temperature to operate \cite{metel01}. The semiconductor technology, now-a- days becomes operable in room temperature \cite{usp}. The semiconductor QCA technology is adopted as the implementation technology for our proposed QCA designs. The molecular technology, not yet implemented, is a single molecule implementation technology. This technology is highly promising due to its highly symmetric QCA cell structure, very high switching speeds, extremely high device density, operation at room temperature and even the possibility of mass-production by means of self-assembly. Magnetic QCA (MQCA) is based on the interaction between the nanoparticles of the magnet. The MQCA depends on the quantum mechanical nature of magnetic interactions, which can be operated at room temperature \cite{mag01} \cite{mag02}.
\section{Related Work}\label{related}
\begin{figure}
\centering
\includegraphics[trim =0cm 0cm 0cm 0cm, clip,scale=0.45]{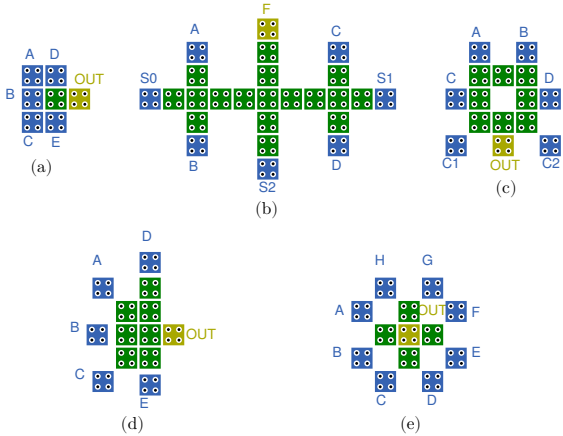}
\caption{Previous reconfigurable gates (a) AOI \cite{momen02} (b) 7-input complex gate \cite{motameni01} (c) RFTG gate \cite{Roohi03} (d) PRC gate \cite{Navi02} (e) Reconfigurable majority gate  \cite{Navi01}}
\label{previousgates}
\end{figure}
Limited attempts have been made to realize reconfigurable designs in QCA \cite{motameni01, momen02, Navi01, Navi02, Roohi03}. An And-OR-Inverter (AOI) logic is proposed in \cite{momen02} where by fixing one or more input various logic functions can be implemented such as OR-AND, NAND-OR, NOR-NAND etc. which is shown in Figure \ref{previousgates}(a). In \cite{motameni01}, a complex 7-input QCA configurable gate (out of 7 inputs, 3 inputs are used as control inputs) is proposed by cascading three 3-input majority voter as shown in Figure \ref{previousgates}(b). The sum of product, product of sum, four-input AND, four-input OR logic etc. can be constructed by fixing control inputs to ``0" or ``1". In \cite{Roohi03}, under the presence of different QCA  defects, a symmetric configurable fault tolerant reconfigurable gate (RFTG) has been proposed as shown in Figure \ref{previousgates}(c). A novel configurable QCA design is proposed in \cite{Navi02}. The design (Figure \ref{previousgates}(d)) is capable of realizing various logic functions such as 2-input OR, 3-input AND, 2-input OR, 3-input OR etc. A reconfigurable majority gate (Figure \ref{previousgates}(e)) is presented in \cite{Navi01}. It is apparent that most of the previous work is limited to implement small logic functions only. Configurable memory structures in QCA are yet to be explored.
\section{Proposed level triggered configurable flip-flop}\label{model}
The main advantage of realizing configurable hardware is low device cost and efficient utilization of device area.  Till date, all the existing configurable designs in QCA strictly follow design rules mentioned below as in \cite{Navi02}:
\begin{itemize}
\item Fixing input cells (control inputs) to logic ``0" or ``1" to produce different functions.  
\item Displacing some of the input cells to change the distance between driver cells of the device to generate different functionality.
\end{itemize}
\par The first method is adopted to realize the proposed configurable designs as mentioned in \cite{Navi02}. To realize a configurable flip-flop, we first choose a JK flip-flop. The characteristic equation of JK flip-flop is $Q(t+1) = J\overline{Q} + \overline{K}Q$. To produce D and T flip-flop, either a complemented value of input J or an un-complemented value of input J need to pass to both the inputs of JK flip-flop. The output function of XNOR is $F = A.B + \overline{A}.\overline{B}$. If we choose B as a control input then we can use XNOR function as a simple QCA wire or an inverter. If B is zero then it will act as an inverter otherwise it will act as a simple QCA wire. So, the XNOR function can be used to produce a complemented value as well as an un-complemented value of its inputs controlling one of the inputs to zero or one. Moreover, a multiplexer is the best possible choice to select any one input between two input values. 
\par The QCA representation of the proposed configurable flip-flop (CFF) is shown in Figure \ref{QCARFF}. The proposed CFF has 5 inputs (A, B, C1, C2 and CLK) and two outputs (Q and $\overline{Q}$) where C1, C2 are control inputs. The primary output of the proposed CFF is as follows:
\begin{equation}
\begin{aligned}
\label{pe}
& Q_{t+1} = \lbrace A.\overline{Q_{t}} + \\
& (\overline{B.C1 + A.C2. \overline{C1} + \overline{A}. \overline{C1}. \overline{C2})}Q_{t} \rbrace CLK +  \overline{CLK}. Q_{t} 
\end{aligned}
\end{equation}
\begin{multline*}
Q_{t+1} = \lbrace A.\overline{Q_{t}} + \lbrace \overline{A}. \overline{B}(C1+C2) + A.\overline{B}(C1+\overline{C2}) + \\ \overline{B}.C1 + \overline{C1}(\overline{A}.C2 + A.\overline{C2})\rbrace Q_{t} \rbrace CLK + \overline{CLK}.Q_{t} 
\end{multline*}
\subsection*{Case 1:}
If C1 = C2 = 0 and CLK=1 then the equation \ref{pe} will be 
\begin{align*} 
Q_{t+1} &= \lbrace A.\overline{Q_{t}} + \lbrace A.\overline{B} + A \rbrace Q_{t} \rbrace  \\
Q_{t+1} &= A.\overline{Q_{t}} + A.Q_{t}  \\   
Q_{t+1} &= A  
\end{align*}
\subsection*{Case 2:}
If C1 = 0, C2 = 1 and CLK = 1 then the equation \ref{pe} will be 
\begin{align*} 
Q_{t+1} &= A.\overline{Q_{t}} + \lbrace \overline{A}.\overline{B} + \overline{A} \rbrace Q_{t} \\
Q_{t+1} &= A.\overline{Q_{t}} + \overline{A}.Q_{t}  
\end{align*}
\subsection*{Case 3:}
If C1 = 1, C2 = X (Don't Care) and CLK = 1 then the equation \ref{pe} will be 
\begin{align*} 
Q_{t+1} &= A.\overline{Q_{t}} + \lbrace \overline{A}.\overline{B} + A.\overline{B} + \overline{B} \rbrace Q_{t} \\
Q_{t+1} &= A.\overline{Q_{t}} + \overline{B}.Q_{t}
\end{align*}
The Table \ref{control} shows the different functionality of the proposed CFF based on control inputs C1 and C2. The CFF will obey the rules of T FF if C1C2 = 01 and if C1C2 = 00 then CFF will behave as D FF. The CFF will behave as JK FF if C1C2 = 1X where X means don't care. The truth table of the proposed CFF is shown in Table \ref{TT}.
\begin{table}[h!]
\centering
\caption{The controlling functionality of proposed CFF}
\scalebox{.84}{
\begin{tabular}{|c|c|c|}\hline 
Input (C1) & Input (C2) & Output (Q)\\\hline
0 & 0 & D FF \\\hline
0 & 1 & T FF\\\hline
1 & X & JK FF\\\hline
\multicolumn{3}{|c|}{X= Don't Care}\\\hline
\end{tabular}}
\label{control}
\end{table}
\begin{figure*}[!t]
\centering
\includegraphics[trim =0cm 0cm 0cm 0cm, clip,scale=0.18]{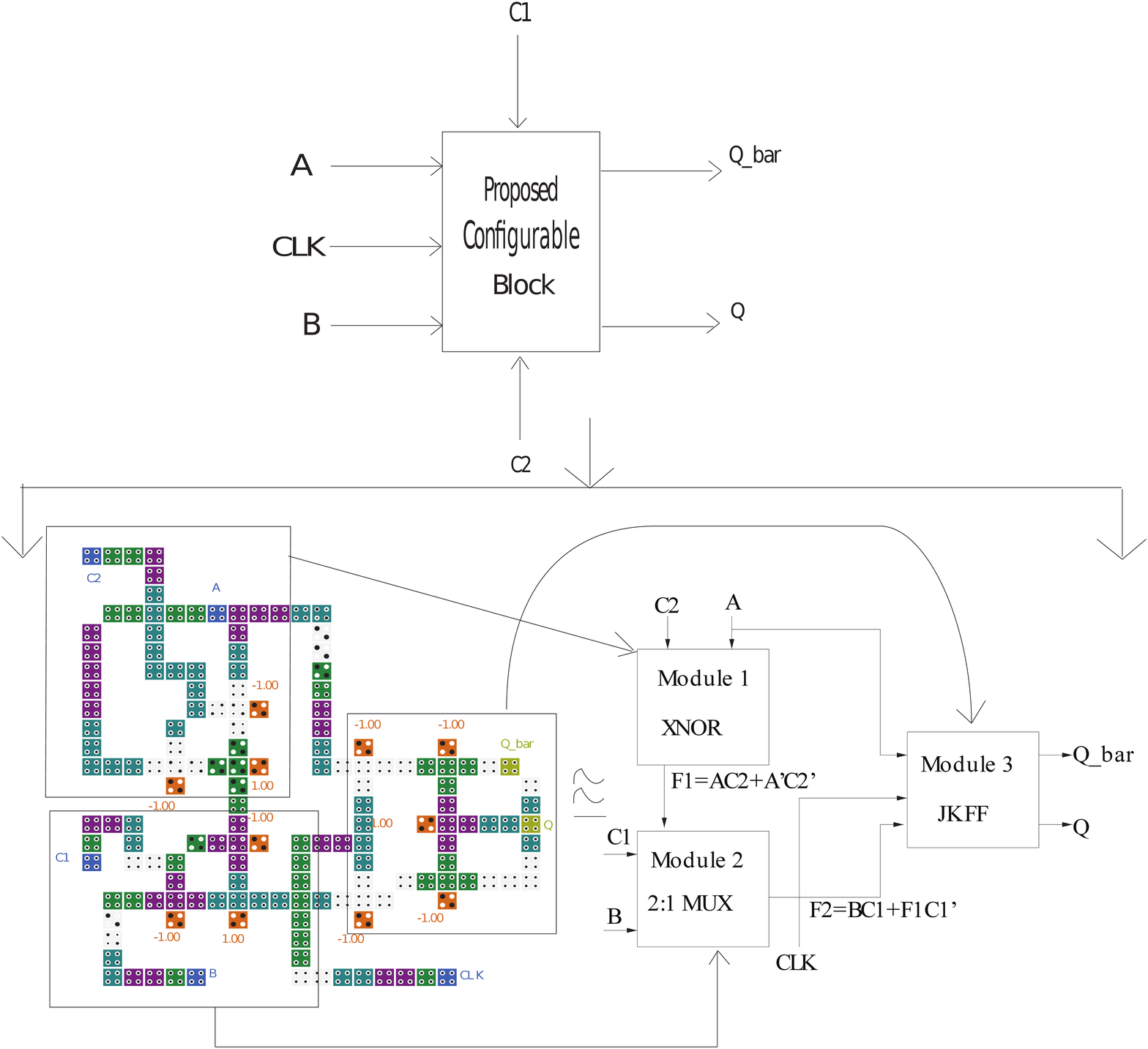}
\caption{The QCA layout of level triggered configurable flip-flop (CFF)}
\label{QCARFF}
\end{figure*} 
\begin{table}[h!]
\centering
\caption{The Truth Table of proposed CFF}
\scalebox{.8}{
\begin{tabular}{|c|c|c|c|c||c|c|}\hline 
A & B & C1 & C2 & CLK & Q & Remark\\\hline
0&0&0&0&1&0& D FF\\\hline
0&0&0&1&1&0& T FF\\\hline
0&0&1&0&1&0 &JK FF\\\hline
0&0&1&1&1&0 &JK FF\\\hline
0&1&0&0&1&0&D FF\\\hline
0&1&0&1&1&0&T FF\\\hline
0&1&1&0&1&0&JK FF\\\hline
0&1&1&1&1&0&JK FF\\\hline
1&0&0&0&1&1&D FF\\\hline
1&0&0&1&1&0&T FF\\\hline
1&0&1&0&1&1&JK FF\\\hline
1&0&1&1&1&1& JK FF\\\hline
1&1&0&0&1&1& D FF\\\hline
1&1&0&1&1&0&T FF\\\hline
1&1&1&0&1&1&JK FF\\\hline
1&1&1&1&1&0& JK FF\\\hline
\end{tabular}}
\label{TT}
\end{table}
\subsubsection*{Module 1 (XNOR)} The function produced by this module is: $F1= A.C2 + \overline{A}.\overline{C2}$. The output of module 1 is passed to module 2 as shown in Figure \ref{QCARFF}. Depending on the value of input C2, module 1 generates a complemented or un-complemented value of input A to module 2. If C2=0 then a complemented value of the input A is passed to module 2 otherwise uncomplemented value of input A is passed to module 2. 
\subsubsection*{Module 2 (2:1 MUX)} The output function of this module is : $F2=B.C1+F1.\overline{C1}$. If C1=1, input B will be selected and module 3 will behave according to JK flip-flop. On the other hand, if C1=0, the module 3 will behave either D flip-flop or T flip-flop depending on the output F1. If $F1= \overline{A}$ then the module 3 will behave as a D flip-flop otherwise it will behave as a T flip-flop.
\subsubsection*{Module 3 (JK FF)} 
\begin{figure*}[!t]
\centering
\subfigure[]{\includegraphics[width=1.5in]{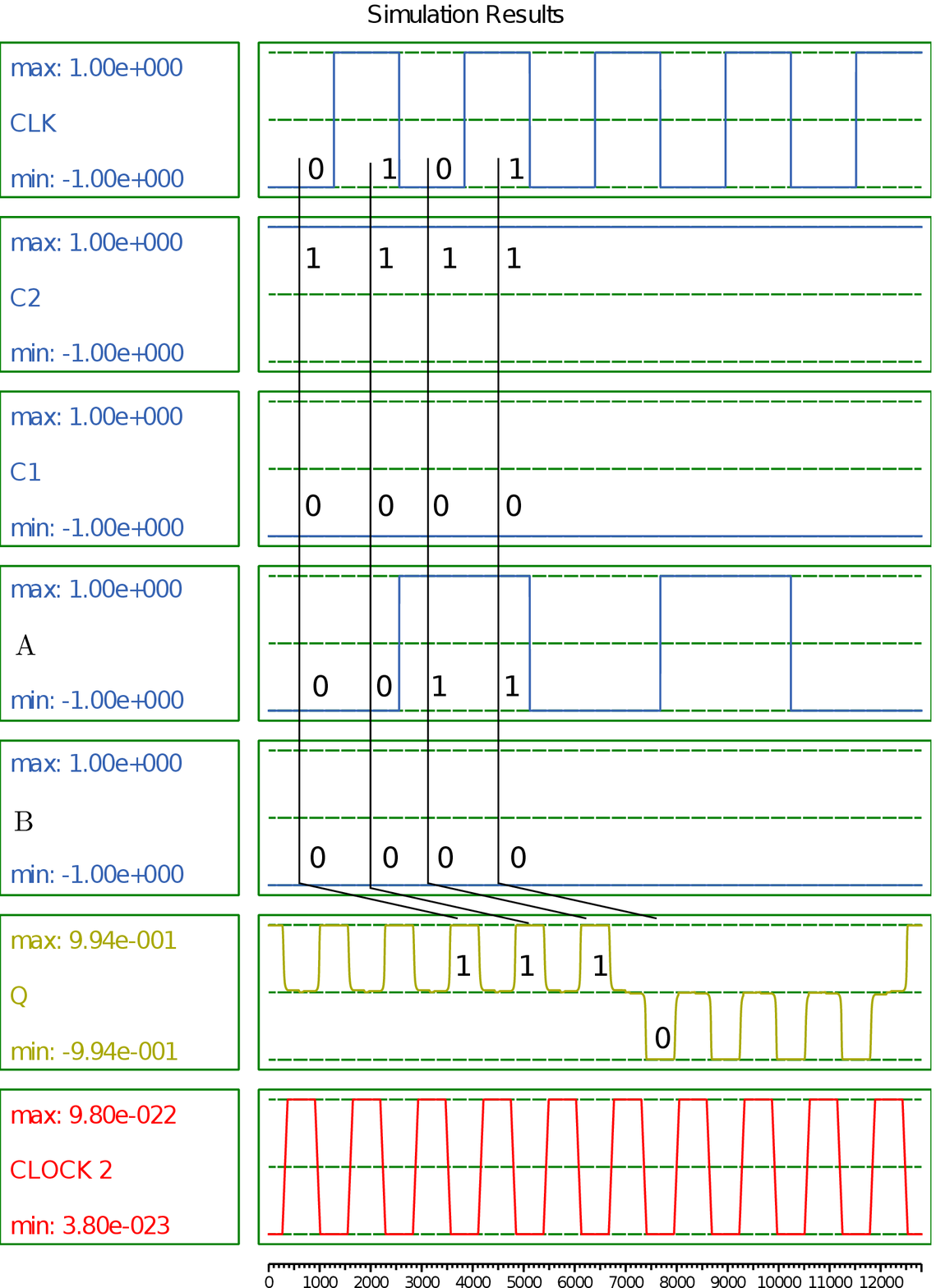}}
\hfil
\subfigure[]{\includegraphics[width=1.5in]{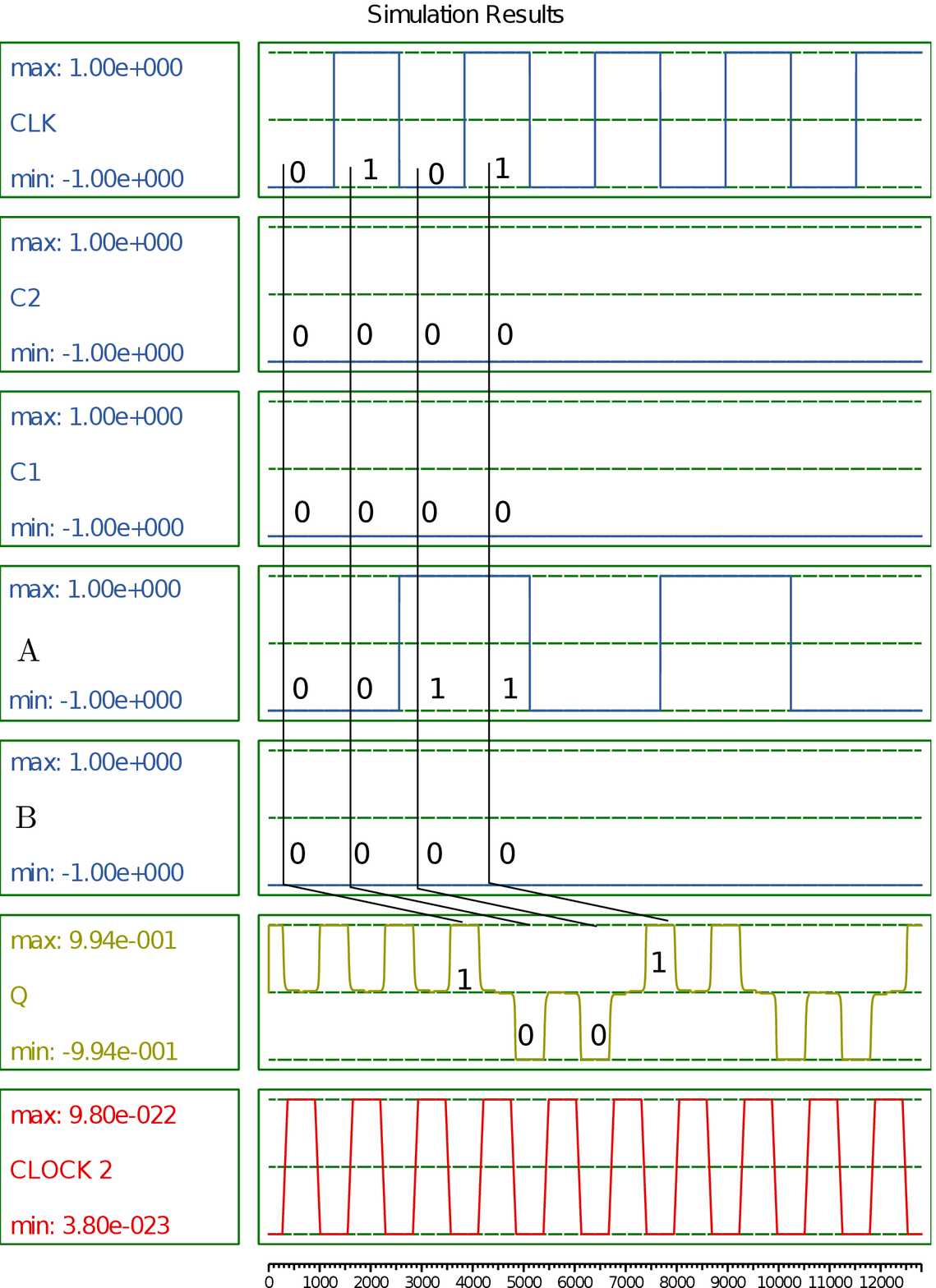}}
\hfil
\subfigure[]{\includegraphics[width=1.5in]{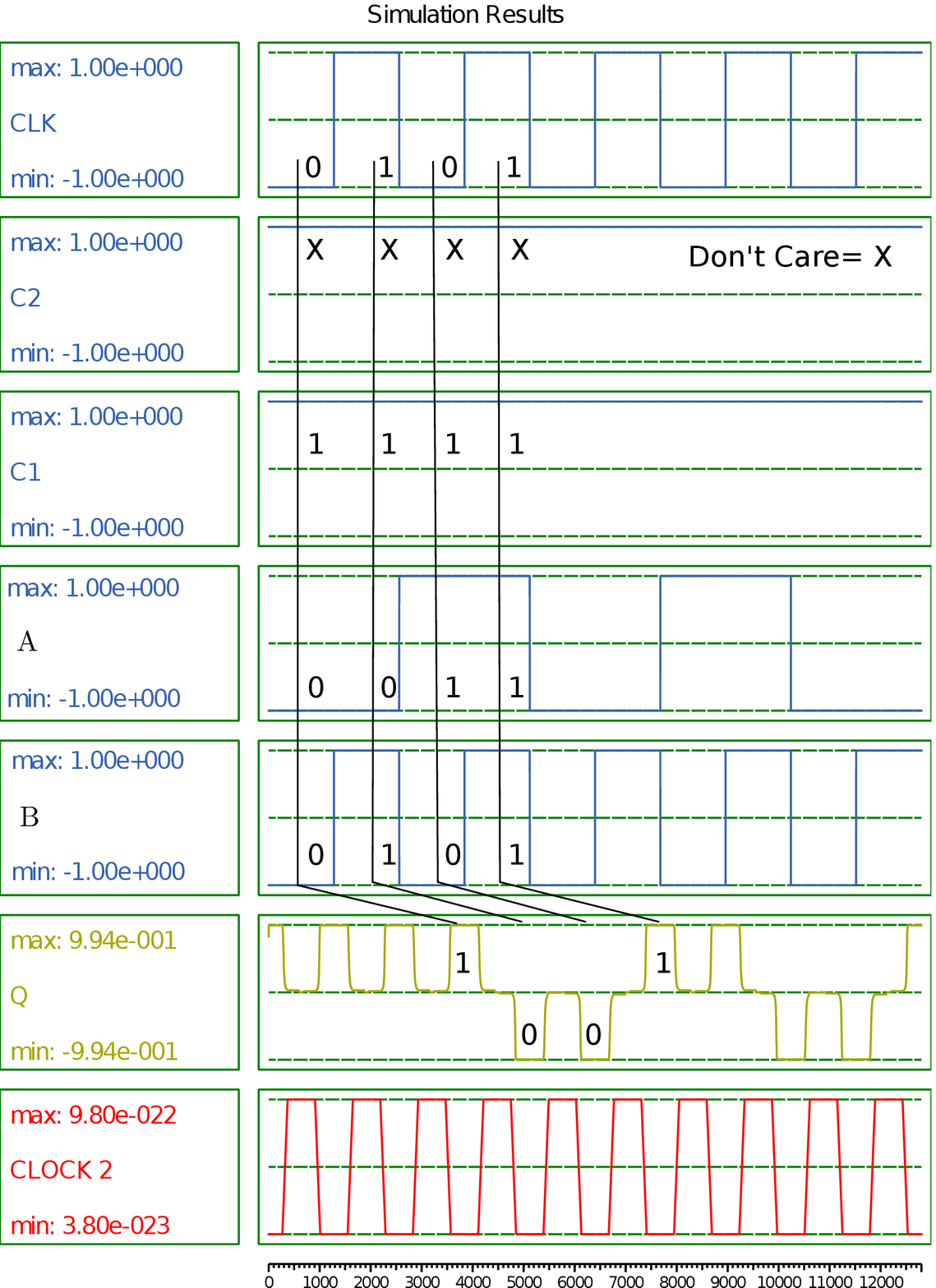}}
\caption{The simulation result of proposed CFF (a) T FF (b) D FF and (c) JK FF}
\label{sim}
\end{figure*}

This module follows the instructions of module 1 and 2. Module 3 behaves as T flip-flop if module 1  produces F1 = A and at the same time module 2 produces F2 = F1 (Figure \ref{sim}(a)). Alternatively, if module 1 produces $F1= \overline{A}$ and at the same time module 2 produces F2 = F1 then module 3 behaves as a D flip-flop. Finally, if module 2 produce F2 = B (module 1: F1= don't care ) then module 3 behaves like a JK flip-flop.
\begin{table}[h!]
\centering
\caption{Performance of proposed CFF design}
\scalebox{.85}{
\begin{tabular}{|c|c|c|c|c|c|}\hline 
Design & Area $\mu m^{2}$ & Cells & Clock & Layer & Configurable \\
       &                  &       & Cycle &       &              \\\hline
\multicolumn{6}{|c|}{The existing level triggered D FF}\\\hline
In \cite{com00} & 0.20 & 104 & 1.25 &Single & No \\\hline
In \cite{com0} & 0.08 & 66 & 1   &Single & No \\\hline
In \cite{FF4} & 0.05& 48 & 1 &Single  & No \\\hline
In \cite{com2} & 0.04& 36 & 1.50 & Single & No\\\hline
\multicolumn{6}{|c|}{The existing level triggered T FF}\\\hline
In \cite{com00} & 0.20 & 108 & 1.25 & Single & No \\\hline
In \cite{com0} & 0.10 & 90 & 1& Single & No \\\hline
In \cite{com2T} & 0.08 & 69 & 1.25& Single & No \\\hline
In \cite{com1T} & 0.06 & 68 & 1.25& Single & No \\\hline
In \cite{comT} & 0.06 & 46 & 1.50 & Single  & No\\\hline
\multicolumn{6}{|c|}{The existing level triggered JK FF}\\\hline
In \cite{comJK} & 0.75 & 415 & 2.50 & Single & No \\\hline
In \cite{com00} & 0.12 & 80 & 1.25 & Single & No \\\hline
In \cite{com0} & 0.10 & 68 & 1 & Single & No\\\hline
\rowcolor{LightCyan}
CFF& 0.20 & 159 & 2.75 & Single & Yes  \\\hline
\end{tabular}}
\label{ComparisionFF}
\end{table}
\par
\begin{figure*}
\centering
\subfigure[]{\includegraphics[trim =0cm 0cm 0cm 0cm, clip,scale=0.28]{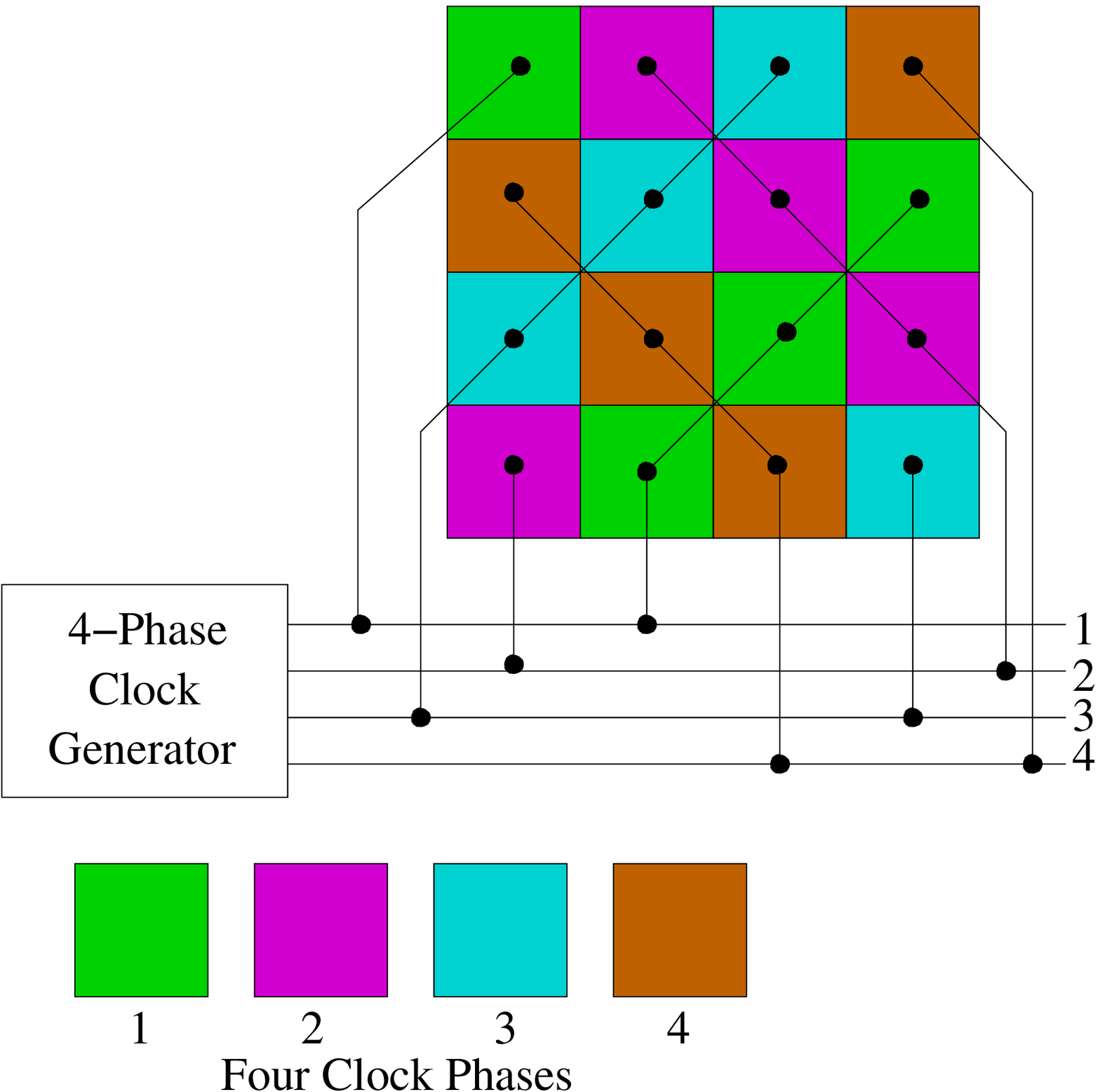}}
\hfil
\subfigure[]{\includegraphics[trim =2cm 13cm 2cm 2cm, clip,scale=0.7]{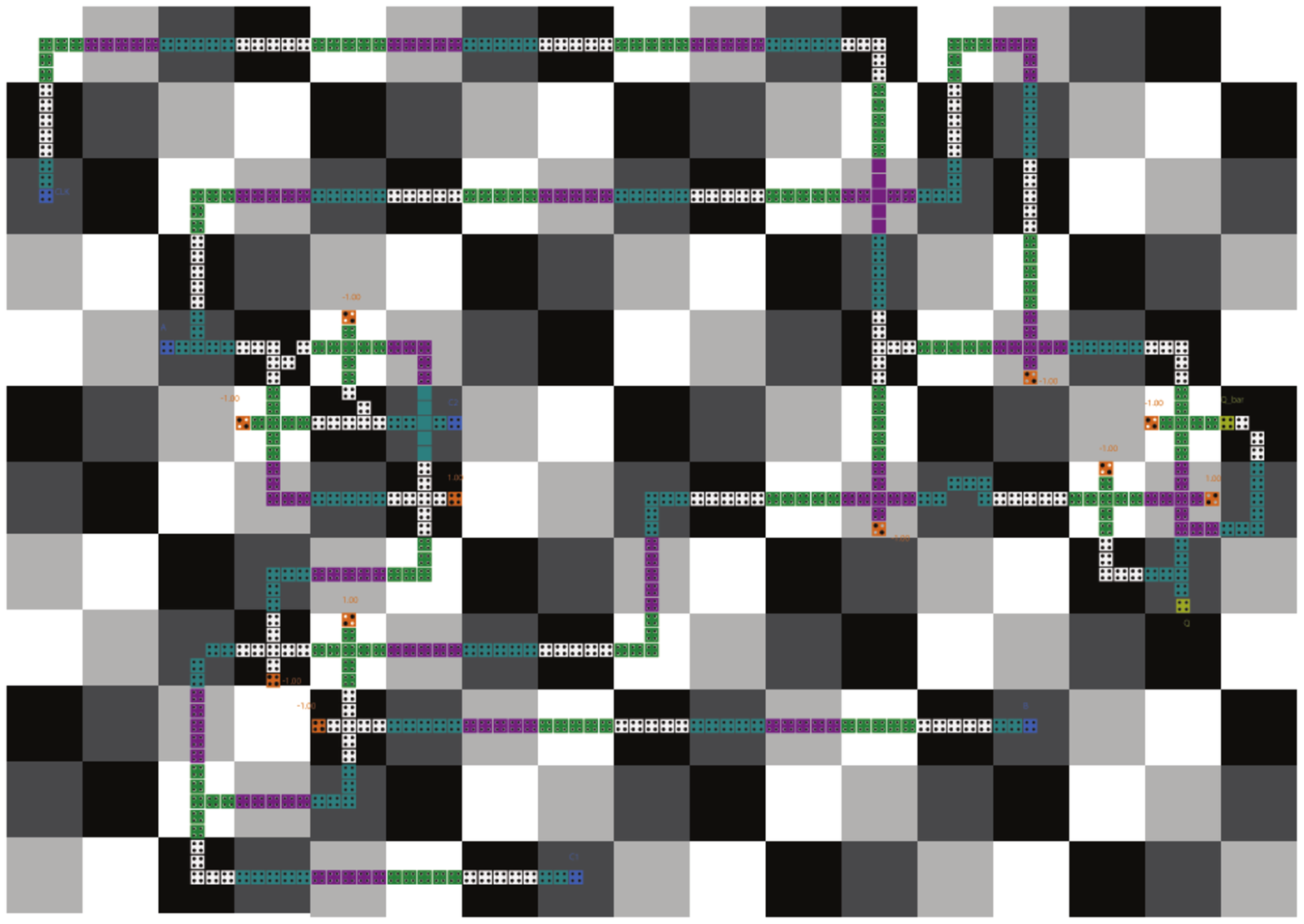}}
\caption{(a) The universal, scalable, efficient (USE) clocking scheme \cite{USE} (b) The CFF layout under USE clocking scheme (CFF-USE)}
\label{USE}
\end{figure*}

 The advantage of configurable memory units lies in its application domain. These can be used to design circuits with architectural similarity eliminating the need to design separate hardware performing a specific function. For example, counters and shift registers can be implemented in a single design with the option to configure the flip-flop according to the need. The biggest advantage of the proposed CFF is that single module can serve as a D, T and JK FF. Table \ref{ComparisionFF} shows the performance of CFF with the existing D, T and JK FF. The proposed CFF consists of $2.75$ delay covering an area of 0.20 $\mu m^{2}$ with 159 QCA cells. Although, the proposed CFF exceeds in all the respect compared to existing designs, but these small overheads can be accepted due to the configurable superiority of CFF over the existing designs. All the previous designs \cite{FF4} \cite{com00} \cite{com0} \cite{com2} \cite{com2T} \cite{com1T} \cite{comT} \cite{comJK} are not configurable in nature and can produce only one function. On the other hand, the CFF can produce three different functions using the same circuit. Moreover, the CFF allows us to present an all in one flip-flop, which can also be perceived as a standard universal design. A universal, scalable, efficient (USE) realistic clock distribution scheme for QCA is proposed in \cite{USE}. The CFF layout under the USE clocking scheme (CFF-USE) is shown in Figure \ref{USE}. The proposed CFF-USE consists of 541 QCA cells covering an area of 1.81 $\mu m^{2}$. 
\begin{figure}
\centering
\subfigure[]{\includegraphics[trim =0cm 0cm 0cm 0cm, clip,scale=0.3]{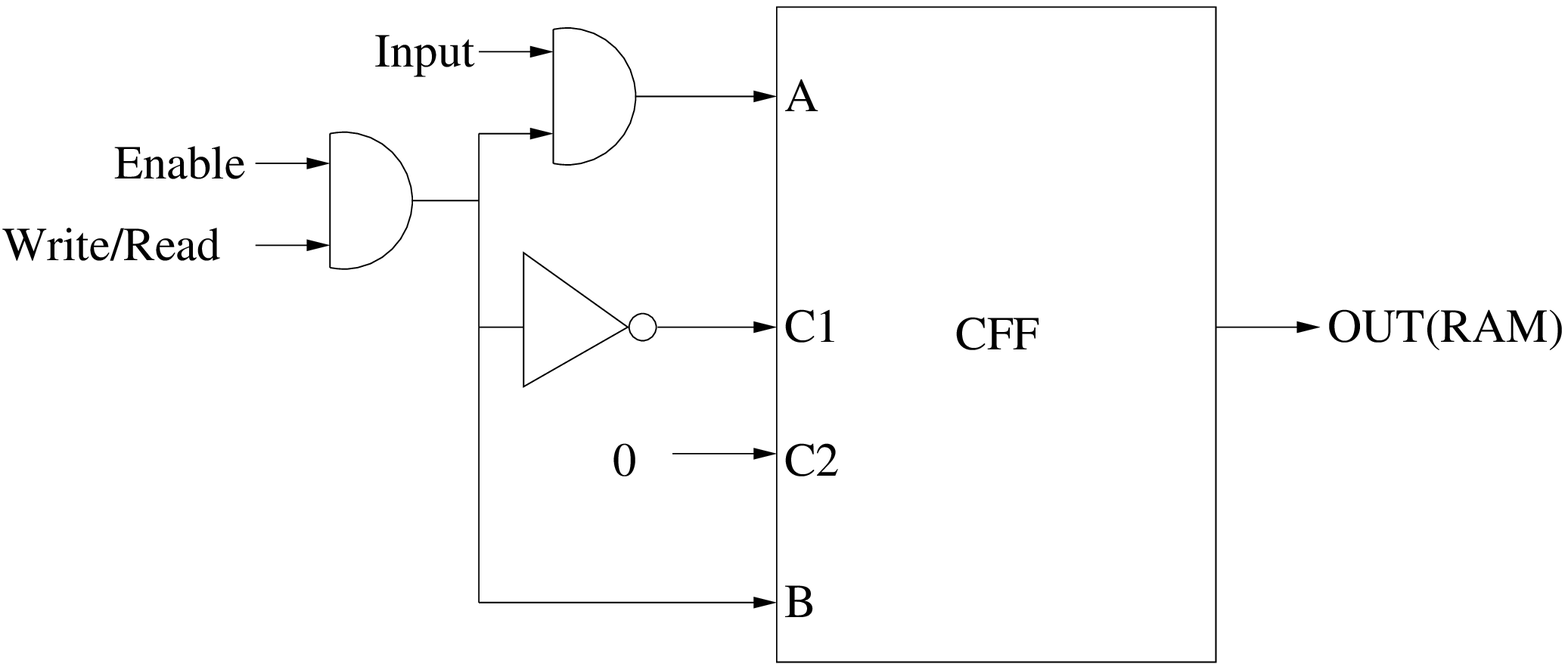}}
\hfil
\subfigure[]{\includegraphics[trim =0cm 0cm 0cm 0cm, clip,scale=0.5]{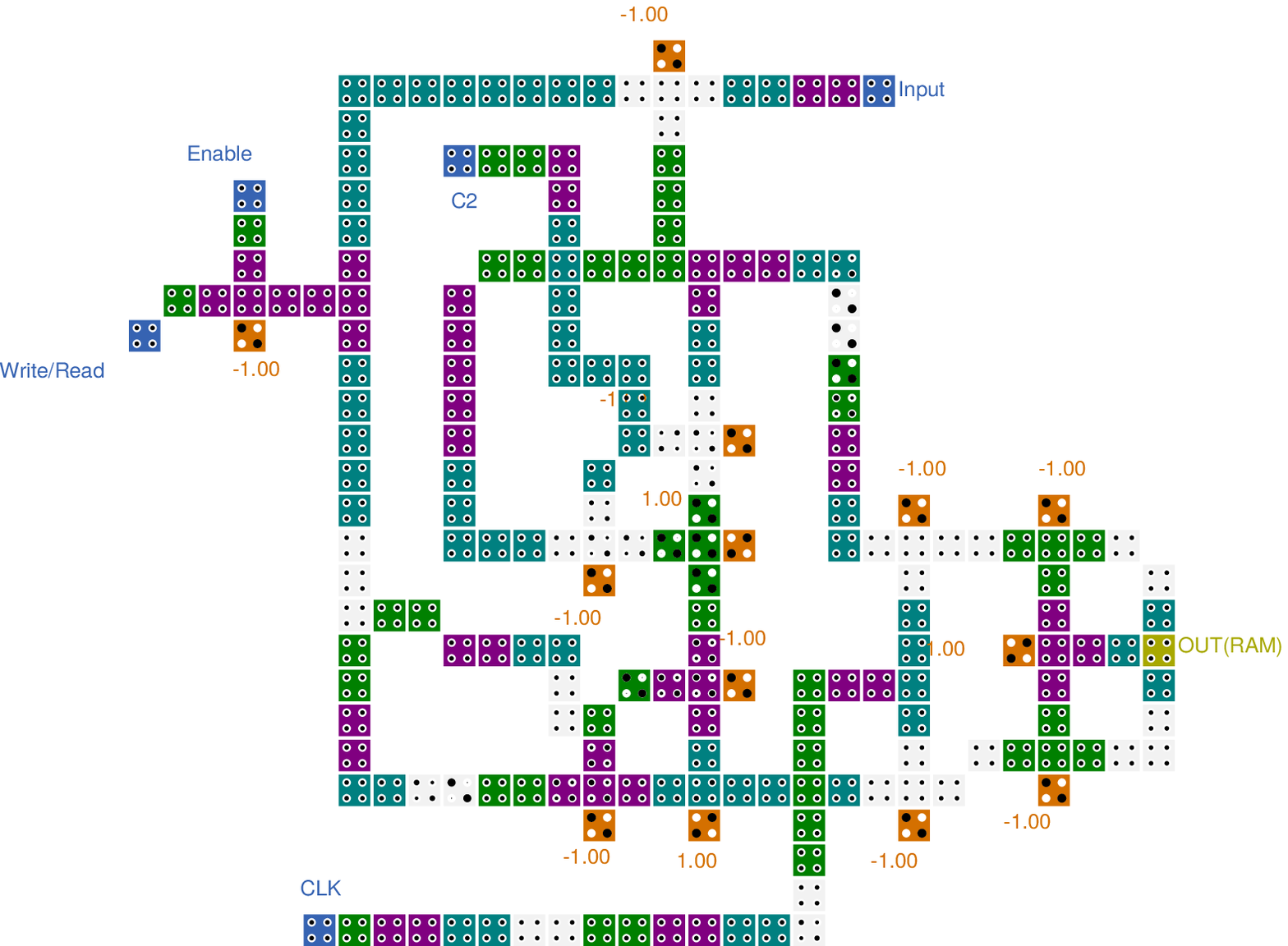}}
\caption{The proposed 1-bit RAM design (a) Logic design (b) QCA representation}
\label{RAM}
\end{figure}
\begin{figure}
\centering
\includegraphics[trim =0cm 0cm 0cm 0cm, clip,scale=0.3]{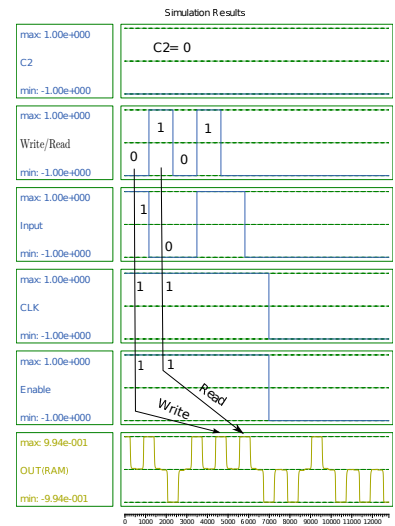}
\caption{The simulation result of proposed RAM}
\label{RAMres}
\end{figure}
\subsection{Memory design using proposed level triggered flip-flop}
In this section, a 1-bit RAM is designed to test the proposed CFF functionality. All the pervious memory (RAM) designs in QCA can be categorized into two groups: line based and loop based. In the loop-based RAM design, four clock zones are utilized to store the previous value but in the line-based RAM design, a QCA wire is used to store the previous value of the output \cite{Angizi2015}. The proposed RAM design also utilizes the property of loop based design which is shown in Figure \ref{RAM}(b). The incoming input value is circulated inside the memory loop if the Enable input is set to 1 and at the same time Write/Read input is set to 0. If the Enable input is set to 1 and Write/Read input is 1 then the current stored value inside the memory loop is fed to the output. In both the cases i.e. read or write, the control input value of C2 is set to zero. The simulation result of the proposed RAM is shown in Figure \ref{RAMres}. The inputs Enable and Write/Read are configured in such a way that they can provide the values needed for C1 and B input (Figure \ref{RAM}(a)). The proposed 1-bit RAM has 209 QCA cells covering an area of 0.31 $\mu m^{2}$. 
\section{Design of  edge configurable flip-flop}\label{DETTT}
The level triggered JK flip-flop is susceptible to noise due to which it leads to race-round condition \cite{FF6}. In order to avoid such unstable phenomenon, edge triggered (falling, rising and dual) flip-flops are extensively studied. There are two well-known schemes in QCA to implement edge triggered flip-flop, the clock pulse generator scheme and the MUX/latch scheme \cite{Xiao2012}. In QCA, the clock pulse generator scheme is explored more than the MUX/latch scheme. Figure \ref{pcg} shows the proposed pulse generator which can produce clock pulses for the falling edge, rising edge as well as dual edges with the help of the two control inputs C3 and C4. To detect a falling edge it takes the help of the previous clock pulse ($CLK_{old}$) and compares it with the current clock pulse (CLK). The $CLK_{old}$ is produced by using four consecutive clock zones (one clock cycle delay) as shown in Figure \ref{pcg}(b). The majority voter representation of the falling edge operation is as follows: 
\begin{figure}
\centering
\subfigure[]{\includegraphics[trim =0cm 0cm 0cm 0cm, clip,scale=0.4]{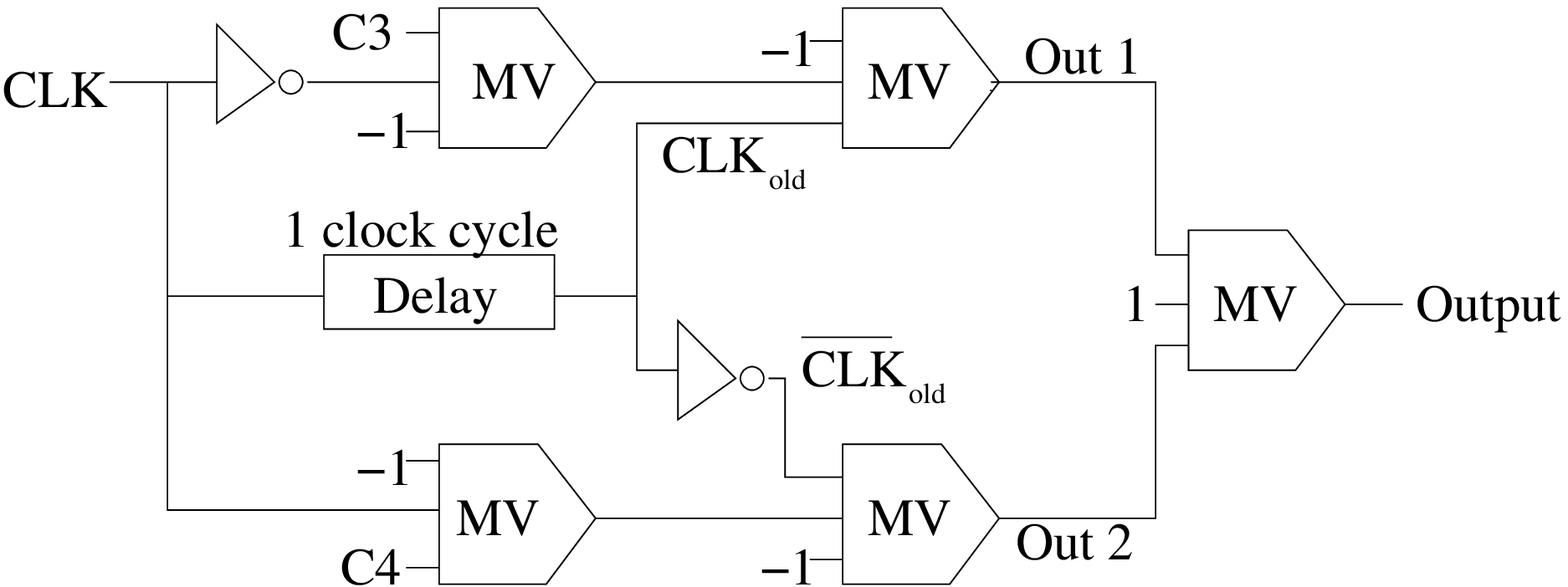}}
\hfil
\subfigure[]{\includegraphics[trim =0cm 0cm 0cm 0cm, clip,scale=0.5]{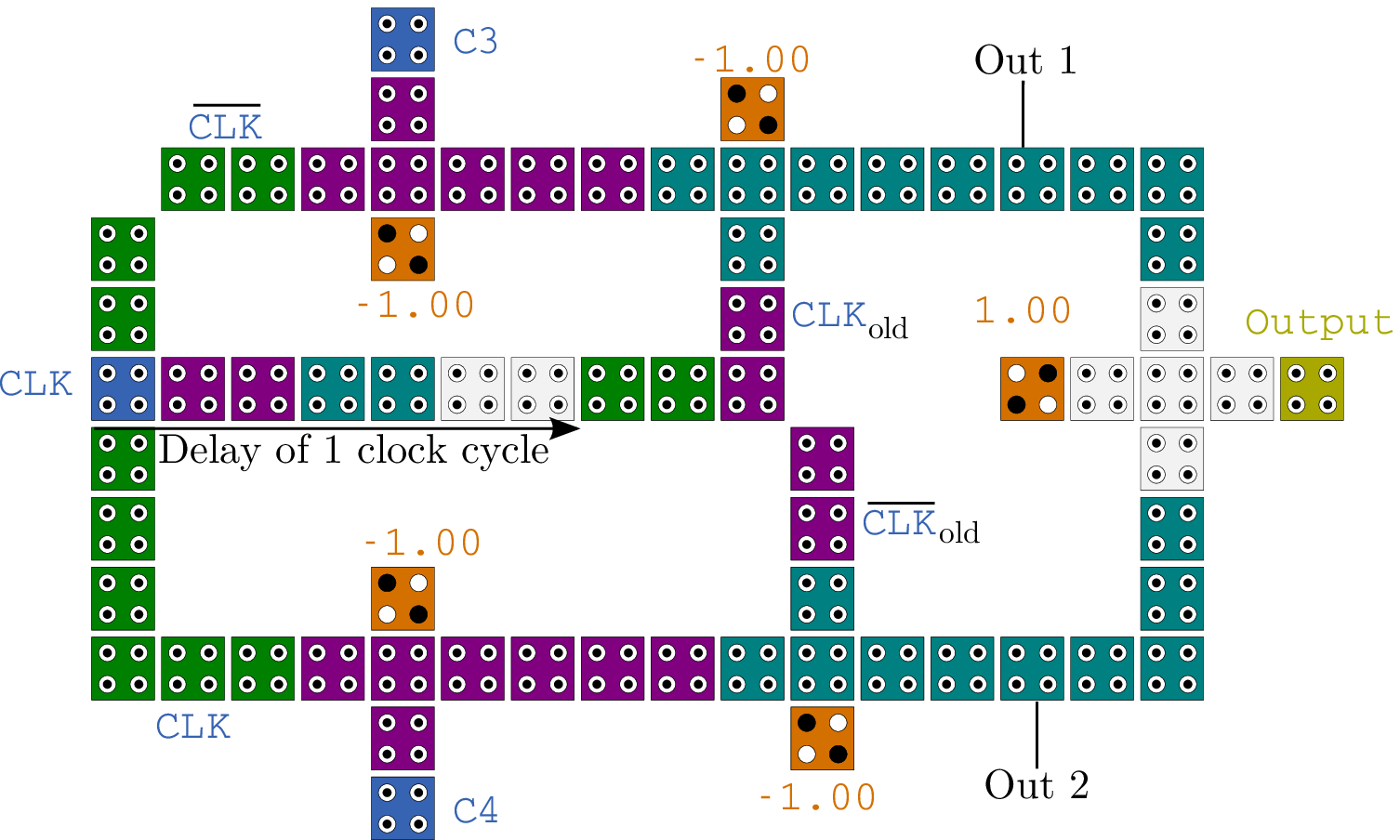}}
\caption{The proposed clock pulse generator (a) Majority voter representation (b) QCA representation}
\label{pcg}
\end{figure}
\begin{figure*}
\centering
\subfigure[]{\includegraphics[trim =0cm 11.5cm 0cm 0cm, clip,scale=.25]{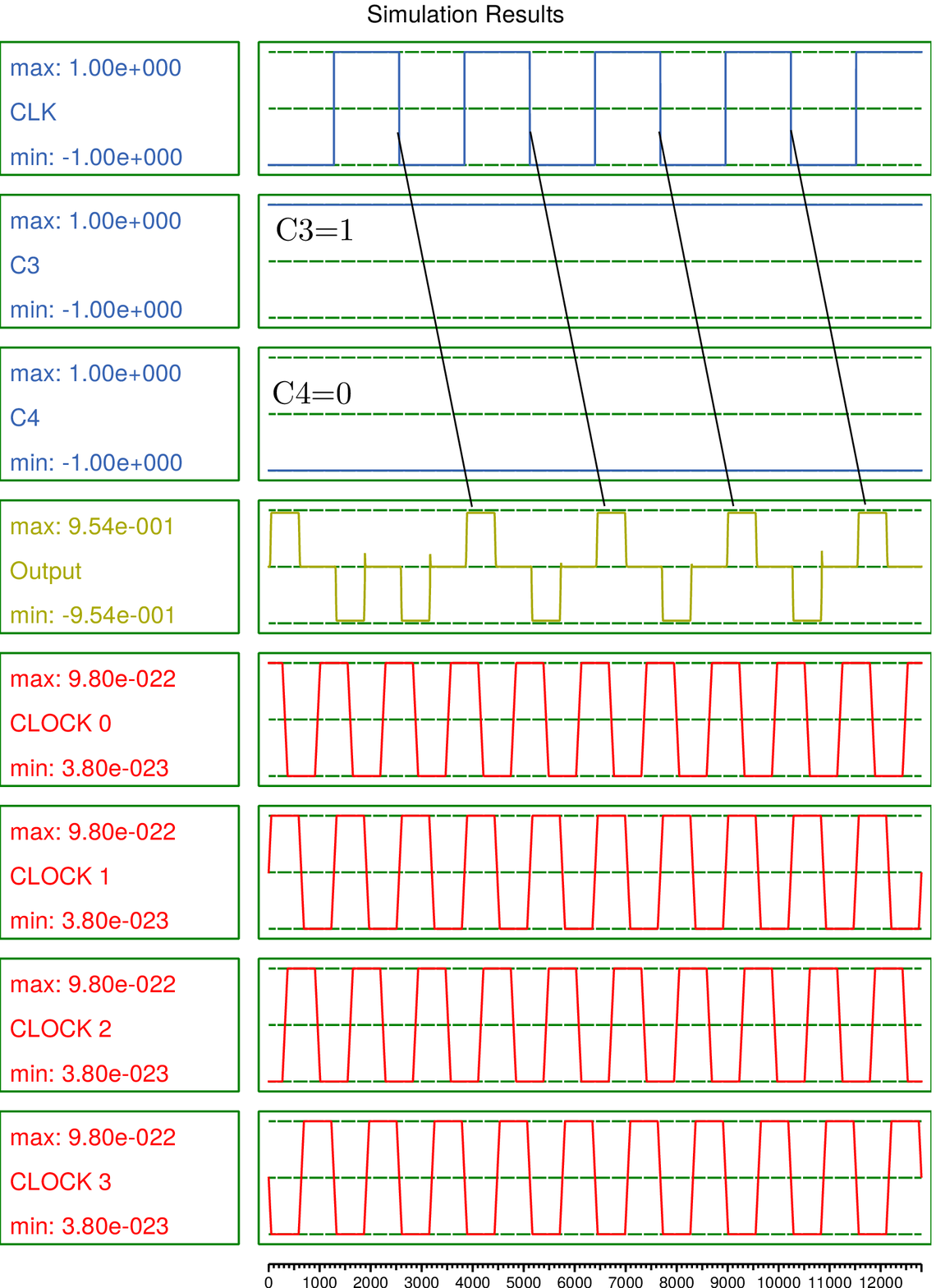}}
\hfil
\subfigure[]{\includegraphics[trim =0cm 11.5cm 0cm 0cm, clip,scale=.25]{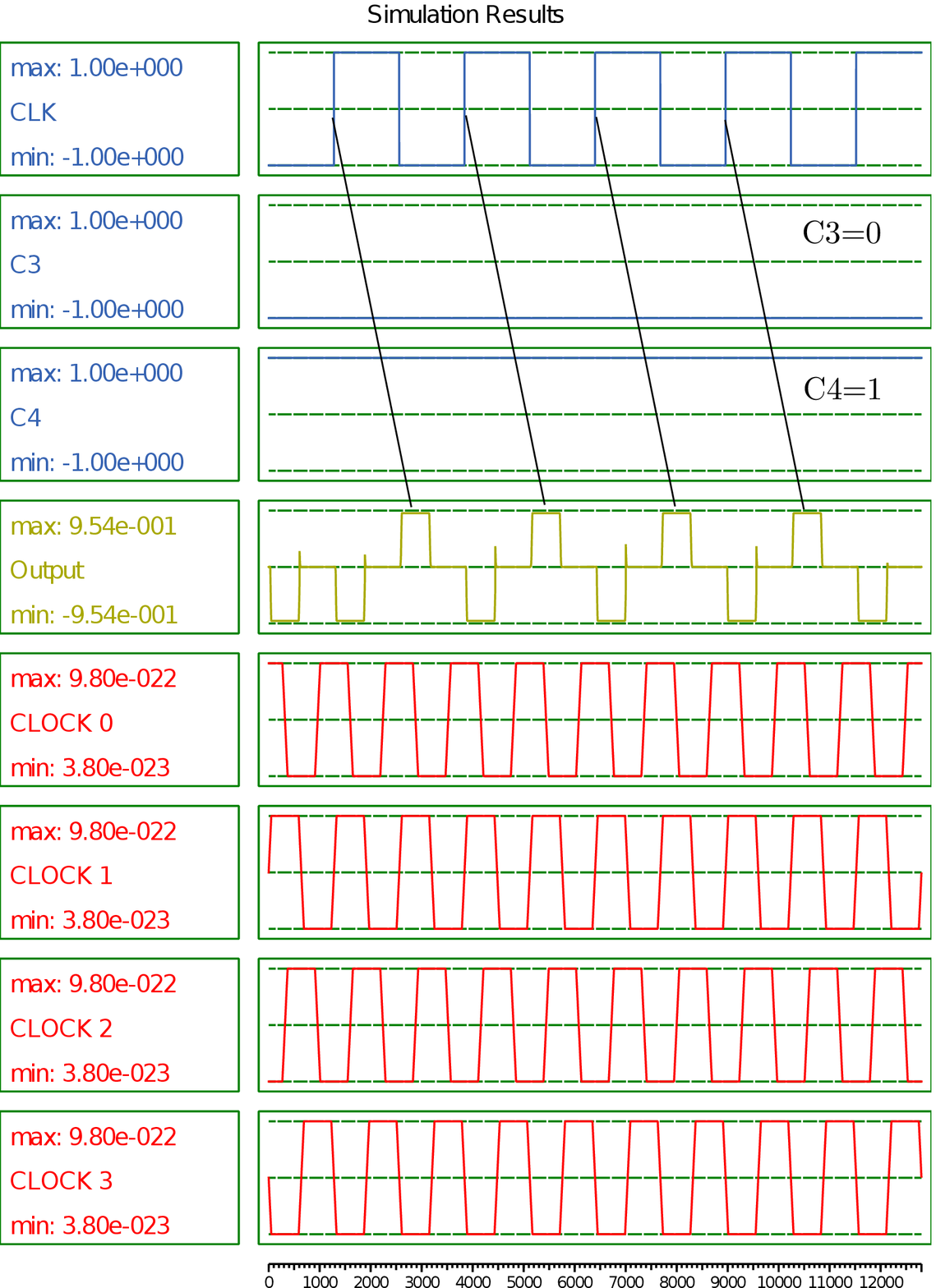}}
\hfil
\subfigure[]{\includegraphics[trim =0cm 11.5cm 0cm 0cm, clip,scale=.25]{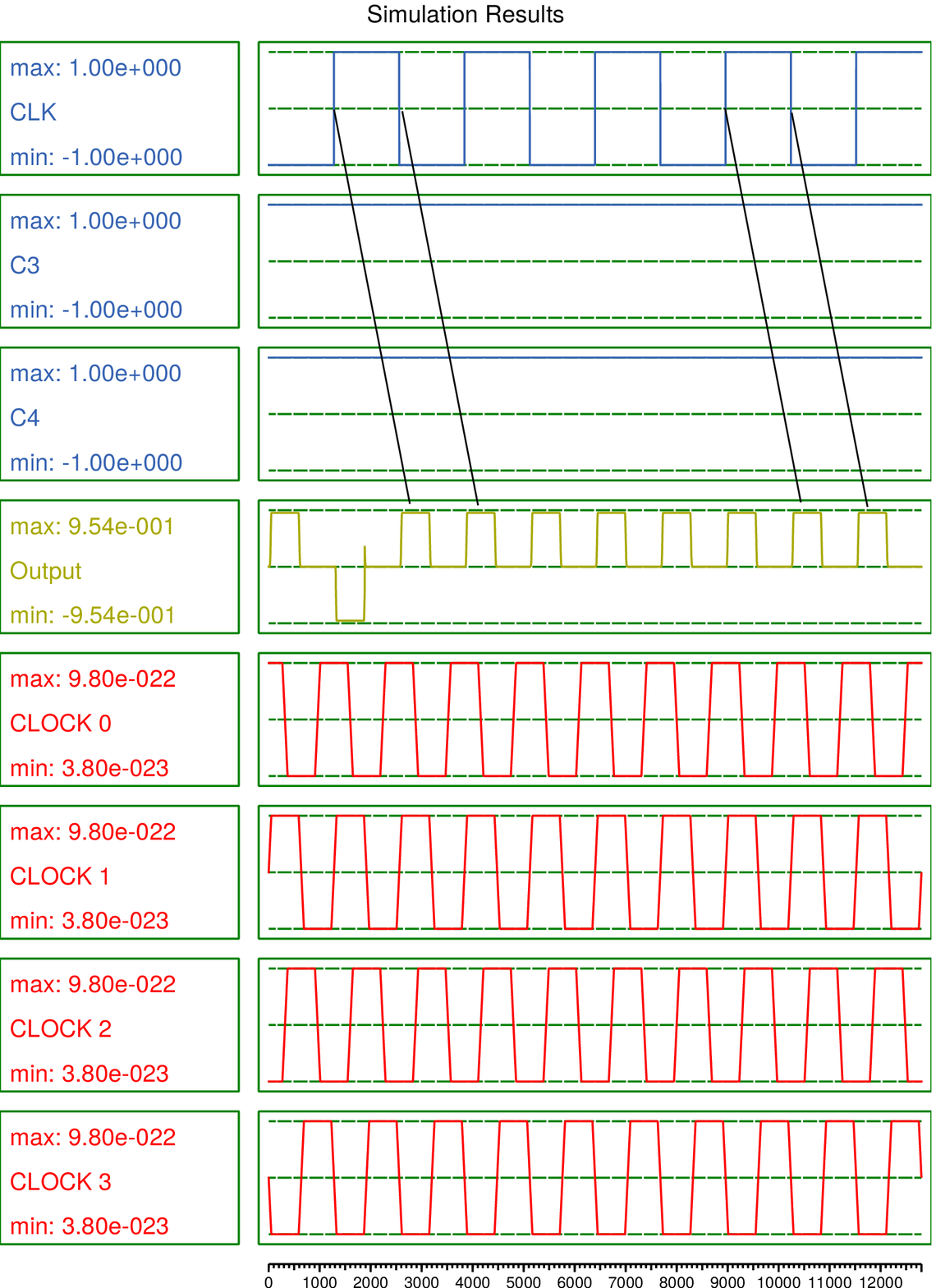}}
\caption{The simulation result of proposed clock pulse generator (a) Falling edge (b) Rising edge and (c) Dual edge}
\label{CPG}
\end{figure*}
\begin{center}
Out1= MV(MV($\overline{CLK}$, C3, -1), $CLK_{old}$, -1)
\end{center}
The value of ($\overline{CLK}$. $CLK_{old}$) will give a resultant boolean value of 1 which triggers a pulse in the output if the control inputs are set to C3=1 and C4=0 (Table \ref{dual}). At the same time, the proposed clock pulse generator can generate rising edge (CLK $\overline{(CLK_{old})}$) which results to 1) if C3=0 and C4=1 (Table \ref{dual}) and hence producing a trigger in the final output. The majority voter representation of the rising edge operation is as follows: 
\begin{center}
Out2= MV(MV(CLK, C4, -1), $\overline{CLK_{old}}$, -1)
\end{center}
The output of Out1 and Out2 (Figure \ref{pcg}(b)) is passed through an OR gate to produce the final output. If both the control inputs (C3=1 and C4=1) are set to 1 then the proposed clock pulse generator can produce pulses for both the edges (rising as well as falling) and hence works as a dual edge triggered clock pulse generator (Table \ref{dual}). The majority voter representation of the dual edge operation is as follows:
\begin{center}
Output=MV(Out1, Out2, 1) 
\end{center}
\begin{table}
\centering
\caption{Operations of clock pulse generator}
\scalebox{.8}{
\begin{tabular}{|c|c|c|c|c|c|c|}\hline 
 C3 & C4 & $CLK_{old}$ & CLK & Out1 & Out2 & Output \\\hline
 \multicolumn{7}{|c|}{The falling edge operation of proposed clock pulse generator}\\\hline
 1  & 0  &   0		  &  0   & 0    & 0   & 0\\\hline
 1  & 0  &   0		  &  1   & 0    & 0   & 0\\\hline 
 1  & 0  &   1		  &  0   & 1    & 0   & 1\\\hline
 1  & 0  &   1		  &  1   & 0    & 0   & 0\\\hline
\multicolumn{7}{|c|}{The rising edge operation of proposed clock pulse generator}\\\hline
 0  & 1  &   0		  &  0   & 0    & 0   & 0\\\hline
 0  & 1  &   0		  &  1   & 0    & 1   & 1\\\hline 
 0  & 1  &   1		  &  0   & 0    & 0   & 0\\\hline
 0  & 1  &   1		  &  1   & 0    & 0   & 0\\\hline
\multicolumn{7}{|c|}{The dual edge operation of proposed clock pulse generator}\\\hline
 1  & 1  &   0		  &  0   & 0    & 0   & 0\\\hline
 1  & 1  &   0		  &  1   & 0    & 1   & 1\\\hline 
 1  & 1  &   1		  &  0   & 1    & 0   & 1\\\hline
 1  & 1  &   1		  &  1   & 0    & 0   & 0\\\hline
\end{tabular}}
\label{dual}
\end{table}
\par The job of the control inputs (C3 and C4) is to activate the pulse generator to get the necessary output pulse. The function of the control inputs are as follows:
\begin{itemize}
\item[1.] If C3=0 and C4=1 then the pulse generator acts as a rising edge triggered pulse generator.
\item[2.] If C3=1 and C4=0 then the pulse generator acts as a falling edge triggered pulse generator.
\item[3.] If C3=1 and C4=1 then the pulse generator acts as a dual edge pulse generator.
\end{itemize}
The simulation waveform of the proposed clock pulse generator is shown in Figure \ref{CPG} which establishes the correctness of the proposed clock pulse generator.
\par
\begin{figure*}
\centering
\subfigure[]{\includegraphics[trim =0cm 0cm 0cm 0cm, clip,scale=.5]{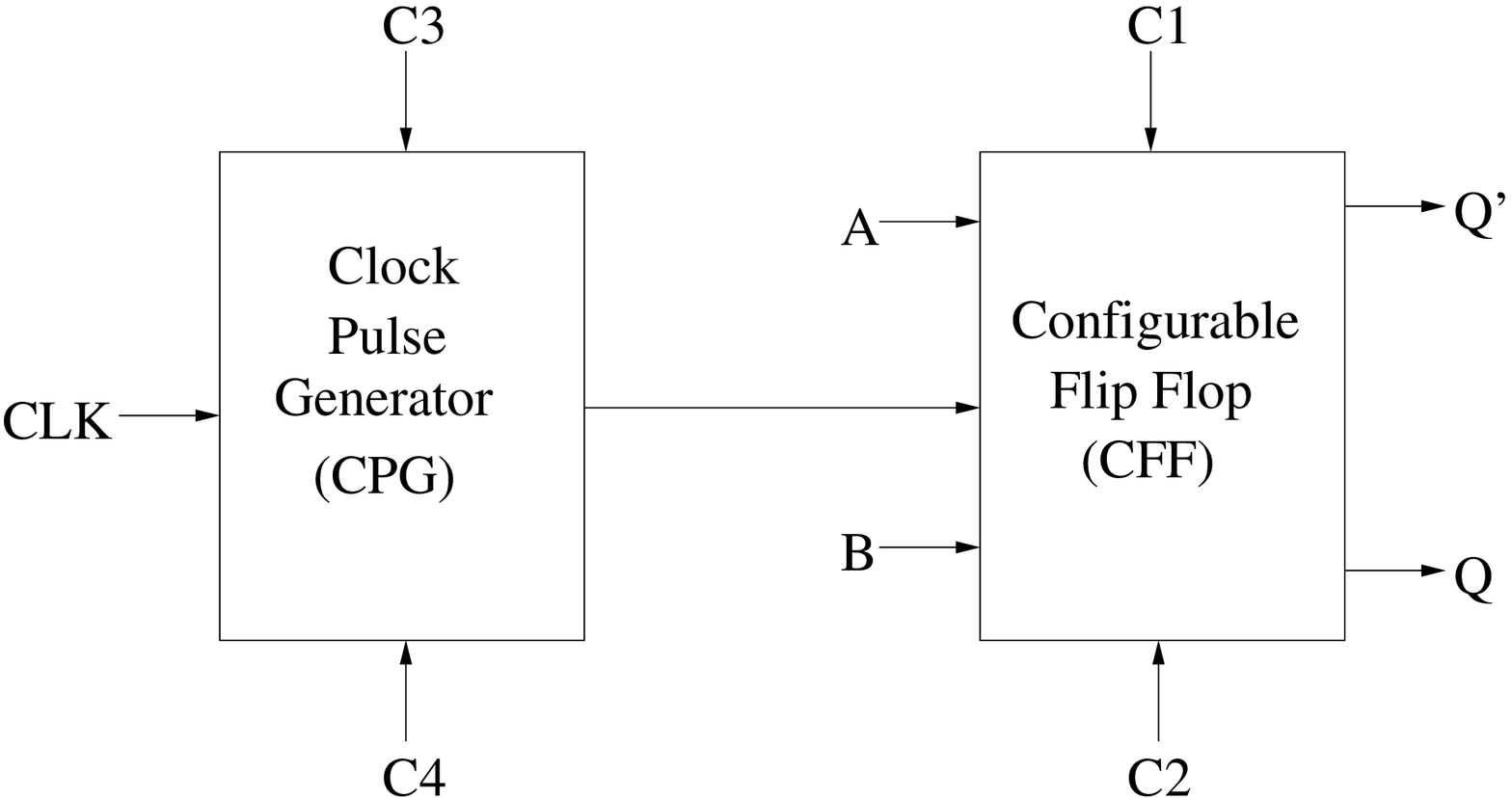}}
\hfill
\subfigure[]{\includegraphics[trim =0cm 0cm 0cm 0cm, clip,scale=.40]{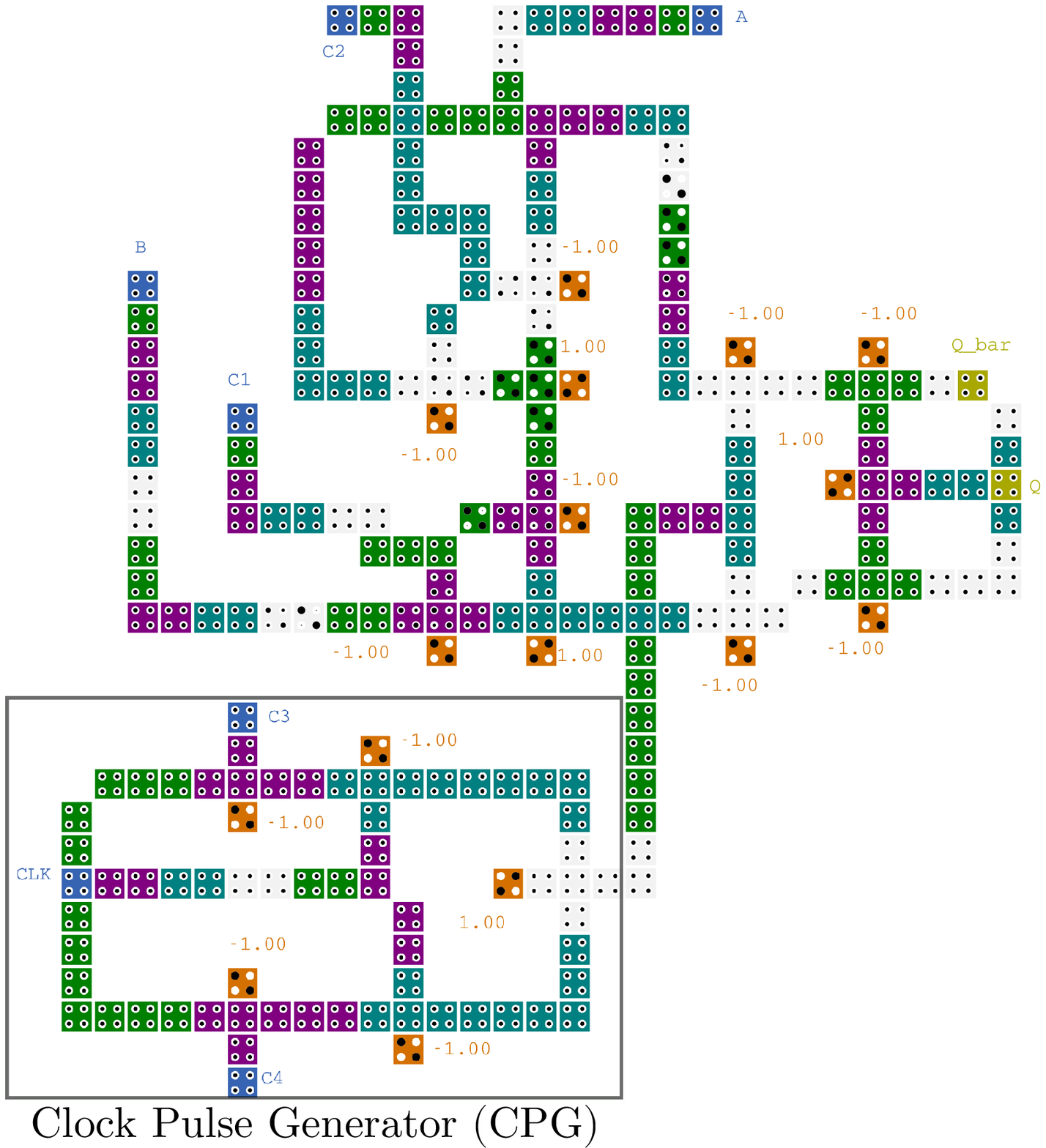}}
\caption{The proposed QCA edge configurable flip-flop (ECFF) (a) Block diagram (b) QCA layout}
\label{DET}
\end{figure*}

\begin{figure*}
\centering
\subfigure[]{\includegraphics[width=1.5in]{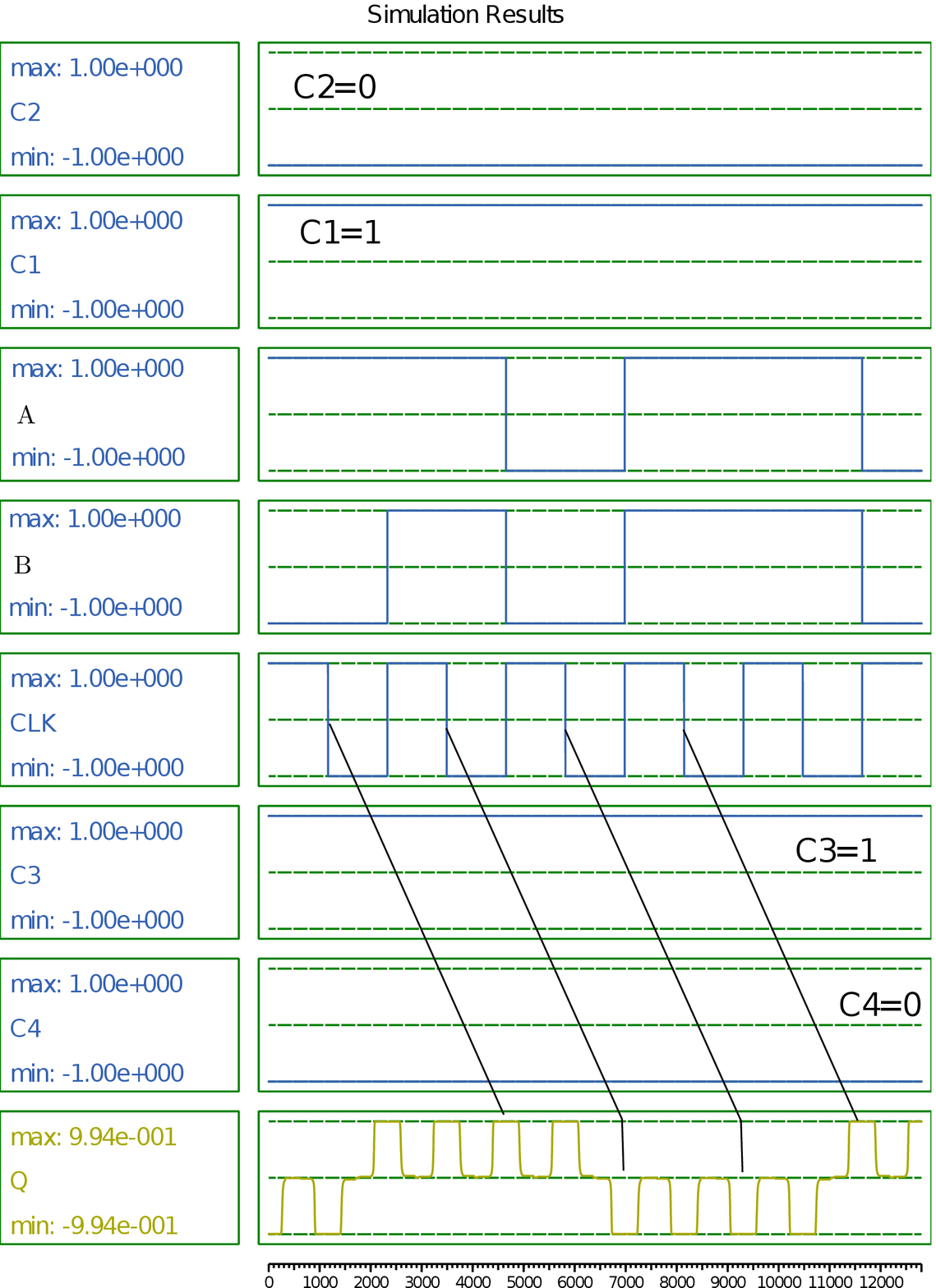}}
\hfil
\subfigure[]{\includegraphics[width=1.5in]{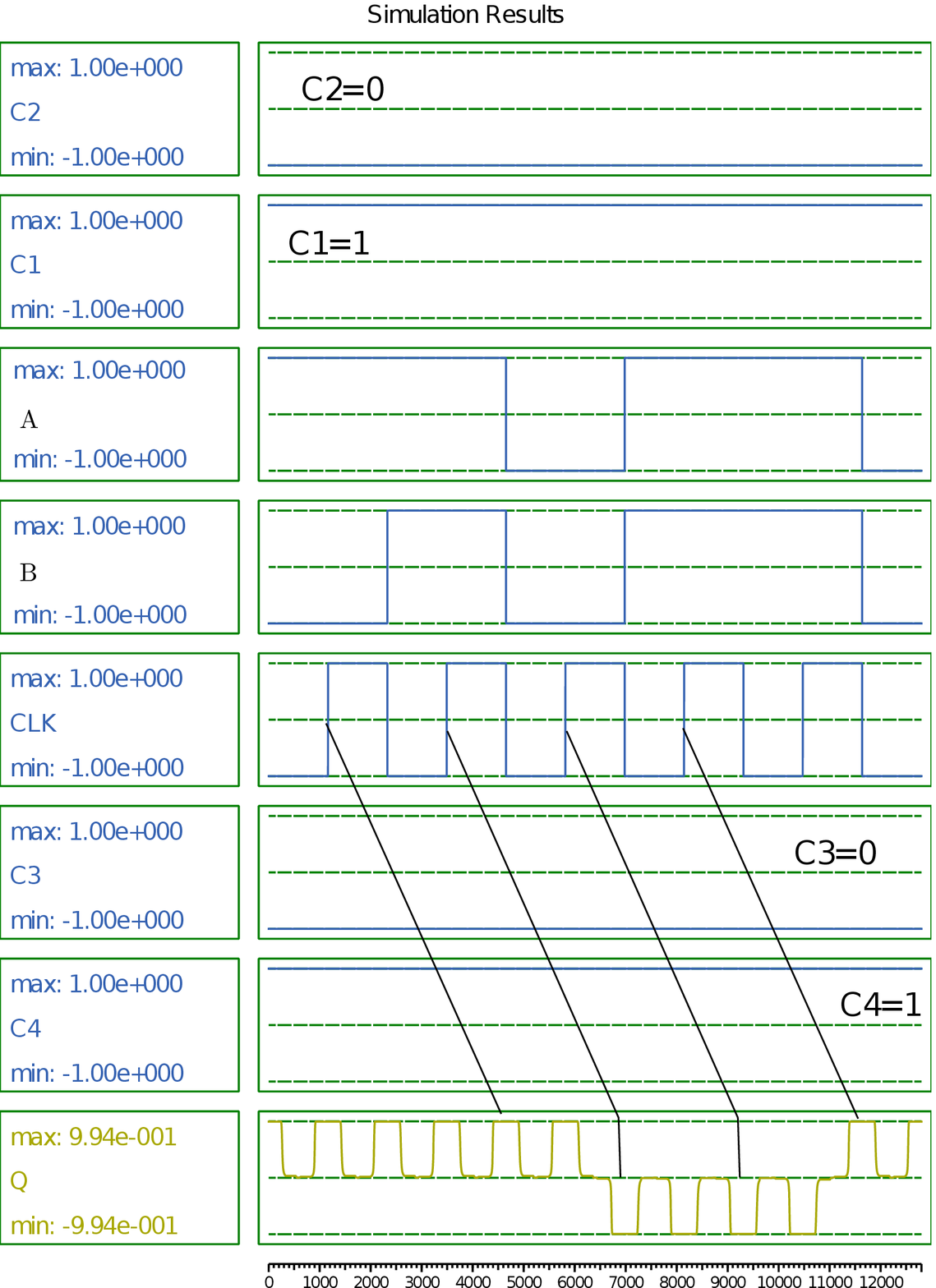}}
\hfil
\subfigure[]{\includegraphics[width=1.5in]{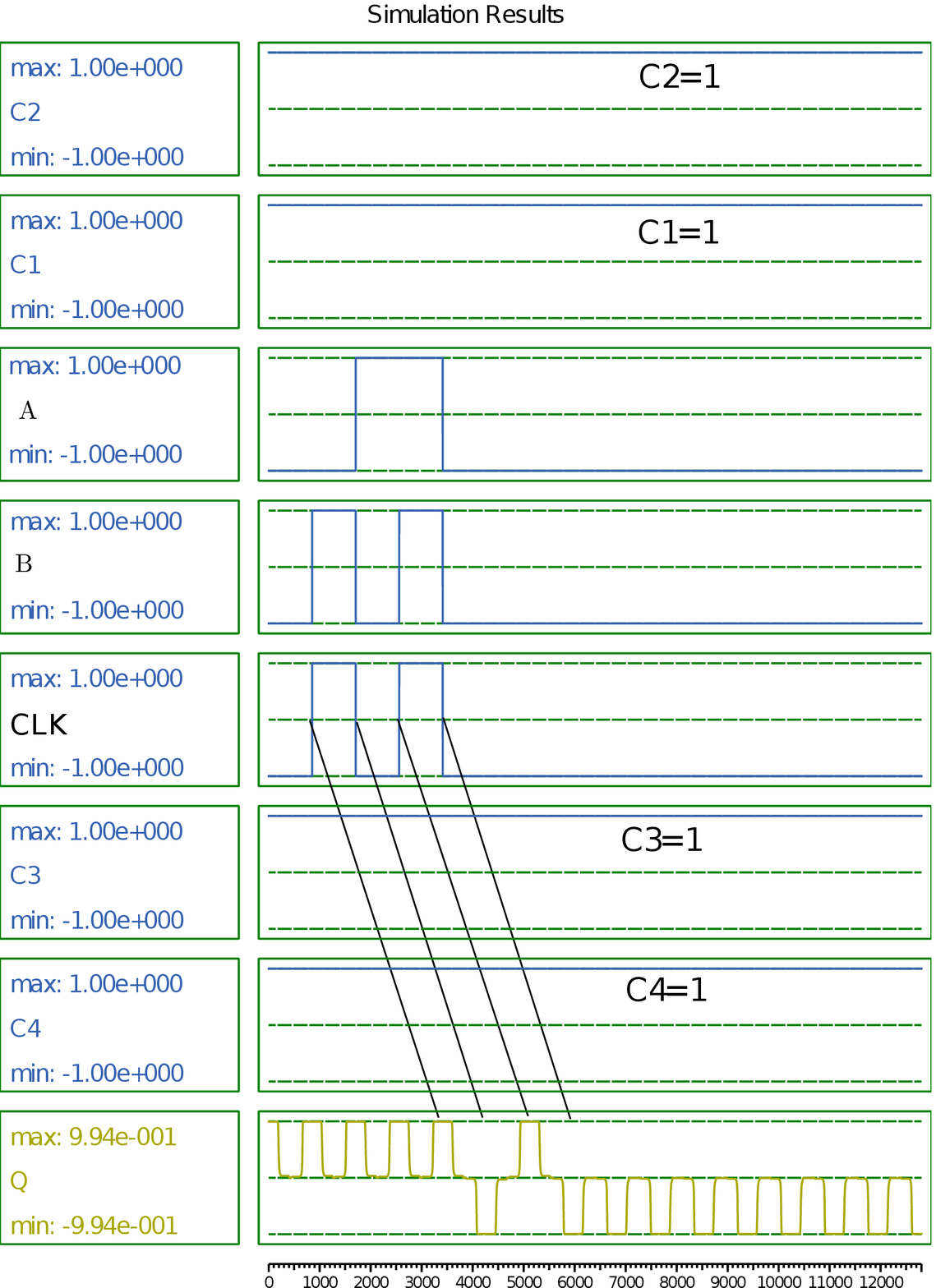}}
\caption{The simulation result of proposed edge configurable JK FF (a) Falling edge JK (b) Rising edge JK and (c) Dual edge JK}
\label{DETSim}
\end{figure*}

 The proposed \textit{edge configurable flip-flop} (ECFF) (Figure \ref{DET}) can be implemented with the help of a clock pulse generator (Figure \ref{pcg}) using CFF as the basic element. Efforts have been made to increase the functionality of ECFF by trying to incorporate a clock pulse generator whose effectiveness is determined by implementing all three types of pulses generated in a single design. The additional design is incorporated within the proposed structure to devise a configurable edge triggered flip-flop. The manifold advantage being that 9 different types of flip-flop designs usually done separately can be incorporated within one. This means the proposed ECFF can be reconfigured to falling edge D/T/JK FF, rising edge D/T/JK FF and dual edge D/T/JK FF using the same circuit. The simulation results of falling/rising/dual edge triggered JK FF is shown in Figure \ref{DETSim}. The performance of the proposed ECFF is shown in Table \ref{PDET}. The existing dual edge triggered flip-flops are compared with the proposed ECFF which can be configured to falling edge D/T/JK FF, rising edge D/T/JK FF and dual edge D/T/JK FF using the same circuit. This means 9 different functions can be produced using the same ECFF whereas the existing flip-flops can produce only one function. 
\begin{table}[h!]
\centering
\caption{Performance of edge configurable flip-flop}
\scalebox{.8}{
\begin{tabular}{|c|c|c|c|c|c|}\hline 
Design & Area  & Cells & Clock& Single & Configurable\\
       & $\mu m^{2}$ &   &Cycle & Layer   &             \\\hline
DET D FF in \cite{FF4} & 0.14 & 120& 3.25 & Yes & No\\\hline
DET D FF in \cite{Xiao2012} & 0.13 & 93 & 1.5 & Yes & No\\\hline
DET D FF in \cite{FF3} & 0.17 & 117& 3  & Yes & No\\\hline
DET T FF in \cite{xiao2013} & 0.32 & 184 & 3& Yes & No\\\hline
DET JK FF in \cite{DETJK} & 0.34 & 197 & 3.25 &Yes & No\\\hline
Proposed ECFF & 0.38 & 242 & 3.75 &Yes & Yes \\\hline
\end{tabular}}
\label{PDET}
\end{table} 
\section{Realization of edge configurable n-bit counter/shift register}\label{counter/register}
\par
\begin{figure}[!h]
\centering
\includegraphics[trim =2cm 17cm 0cm 3.33cm, clip,scale=.4]{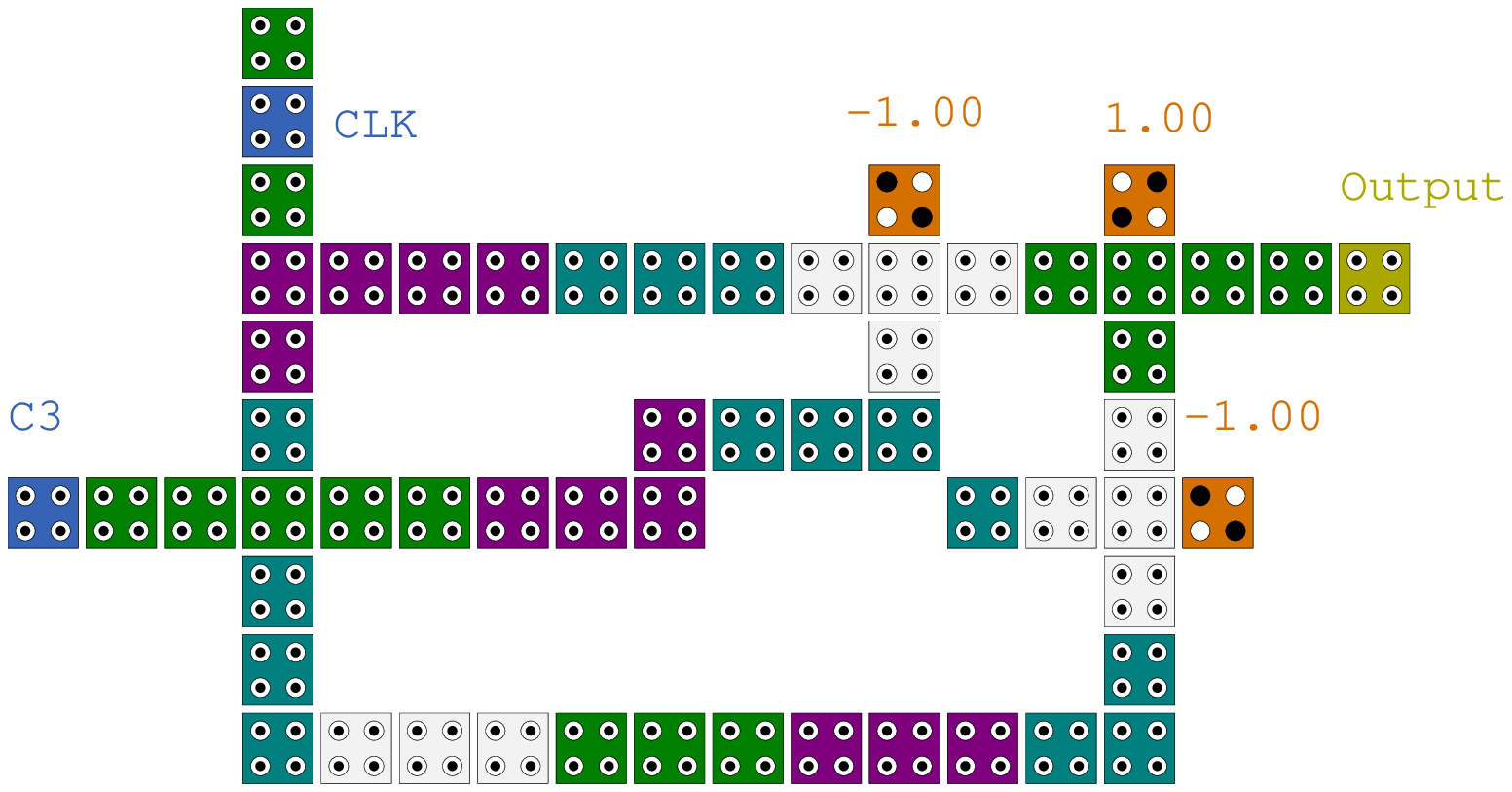}
\caption{The delay control circuit}
\label{delayckt}
\end{figure}

 In this section, the edge configurable flip-flop is used as a basic element to realize the proposed counter/ shift register. In order to synchronize the clock of the proposed configurable counter/shift register one additional delay control circuit (DCC) (Figure \ref{delayckt}) is introduced. The additional delay circuitry is used to delay the clock (fixing C3=0) by one complete cycle so that the clock can be synchronized. The function of the delay circuit is shown in Table \ref{opt}.
\begin{table}
\centering
\caption{The operation table of delay control circuit}
\scalebox{.80}{
\begin{tabular}{|c|c|c|c|}\hline 
C3 & Delay $(CLK_{t-1})$ & Current $(CLK_{t})$& Output \\\hline
0 & 0 & 0 & 0\\\hline
0 & 0 & 1 & 0\\\hline
0 & 1 & 0 & 1\\\hline
0 & 1 & 1 & 1 \\\hline
1 & 0 & 0 & 0\\\hline
1 & 0 & 1 & 1\\\hline
1 & 1 & 0 & 0\\\hline
1 & 1 & 1 & 1\\\hline
\end{tabular}}
\label{opt}
\end{table}
\par
\begin{figure*}
\centering
\includegraphics[trim =0cm 0cm 0cm 0cm, clip,scale=.7]{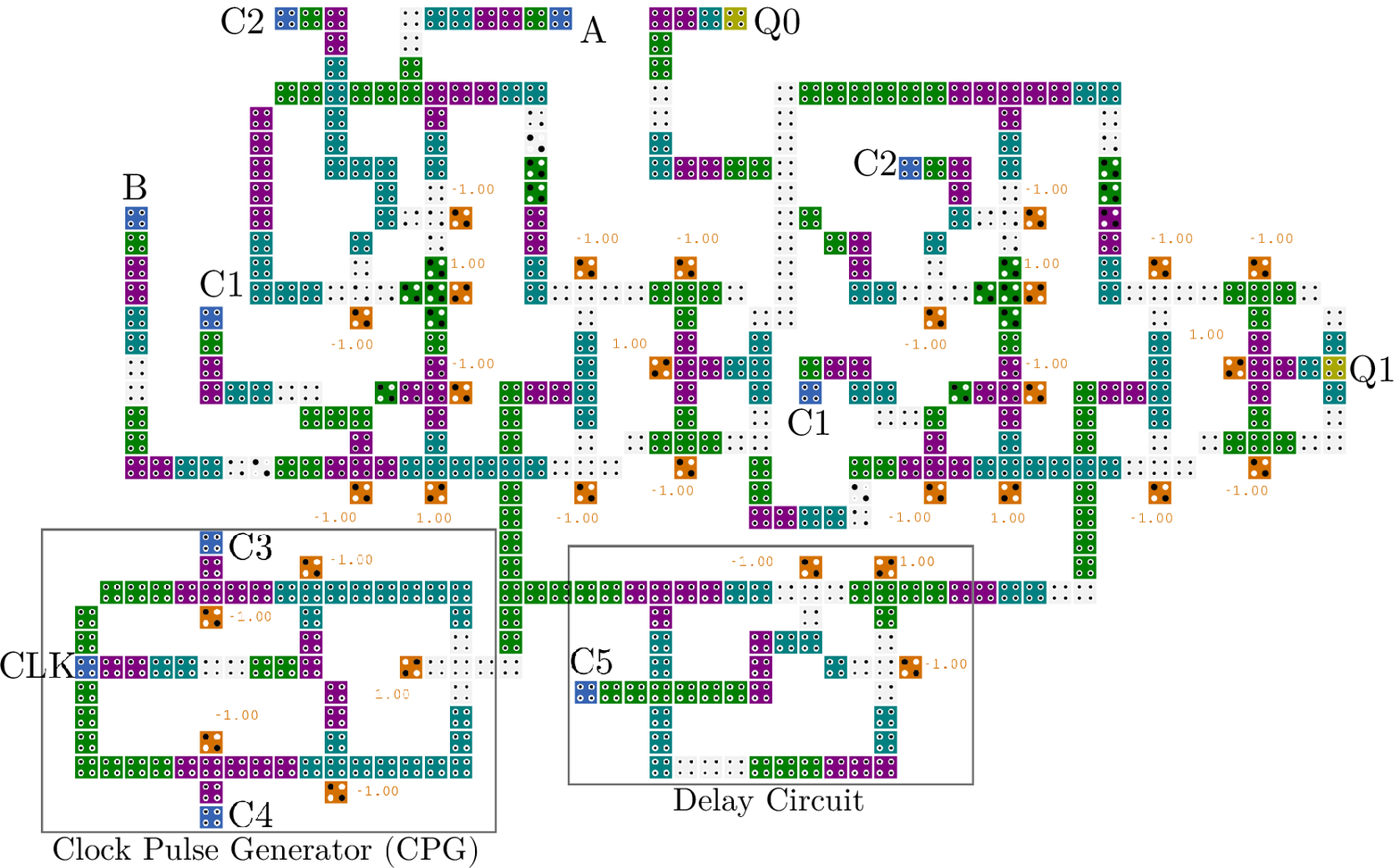}
\caption{The proposed QCA edge configurable 2-bit counter/shift register}
\label{2-bitDET}
\end{figure*}

 The proposed configurable 2-bit counter/ shift register is shown in Figure \ref{2-bitDET}. It can be configured to a shift register or a counter as necessary. The proposed configurable 2-bit counter/ shift register has 7 control inputs (C1, C1, C2, C2, C3, C4, C5), two primary inputs (A, B) and two primary outputs ($Q_{0}$ and $Q_{1}$). The basic unit of the proposed counter/ shift register is the edge configurable flip-flop (CFF). A counter can be constructed using T or JK FF whereas shift register can be constructed using D FF. In the proposed 2-bit counter/ shift register, C1 and C2 inputs are used to adjust the CFF module to the required flip-flop whereas C5 input defines the behaviour of the circuit i.e. if C5=0 then the circuit behaves as a shift register otherwise it behaves as a counter. The two ECFF (configured as a D FF; C1=C1=0, C2=C2=0 and C5=0) are cascaded to design the proposed 2-bit shift register. 
\par
\begin{figure*}
\centering
\includegraphics[trim =0cm 0cm 0cm 0cm, clip,scale=.30]{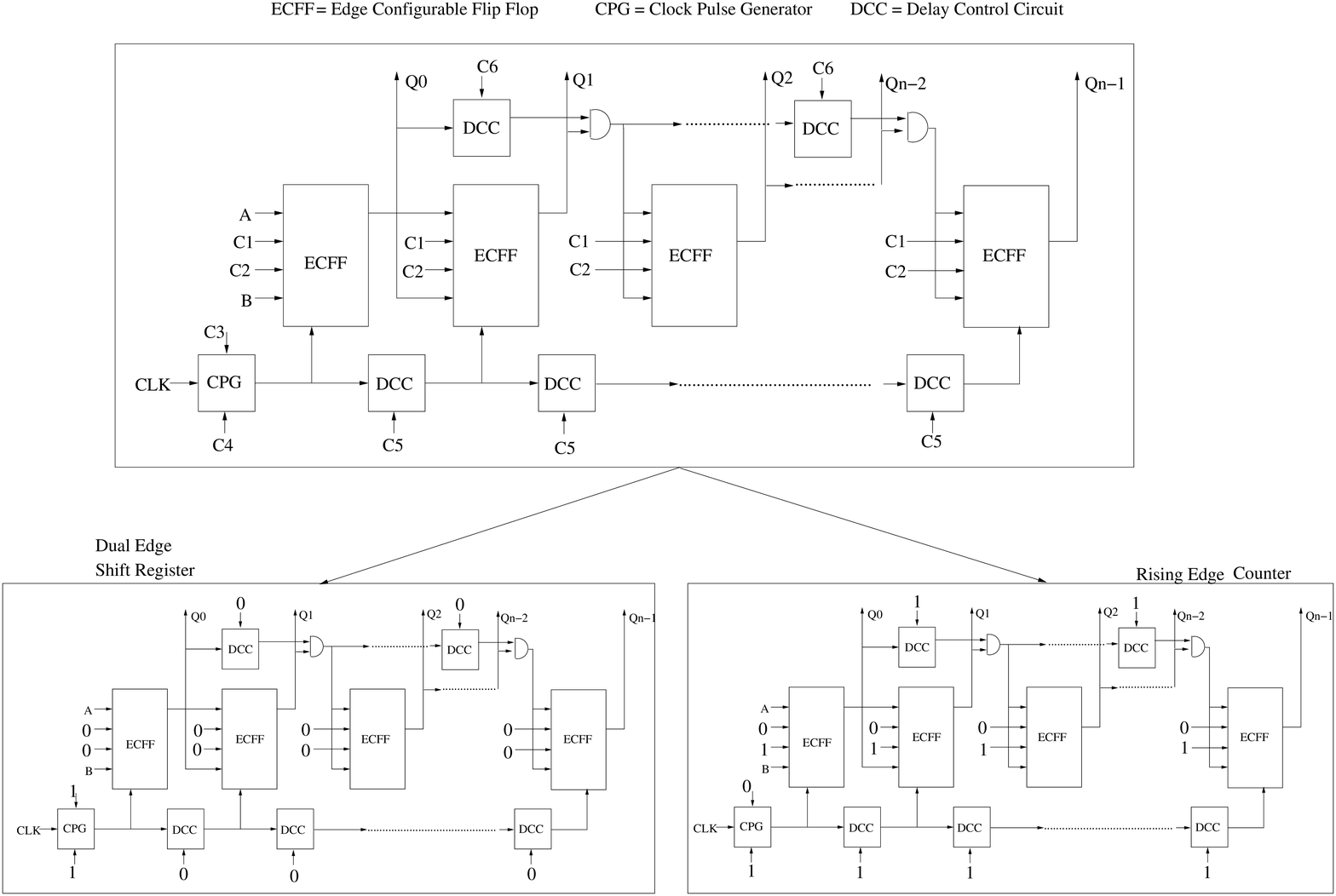}
\caption{The schematic representation of QCA n-bit edge configurable counter/shift register}
\label{schematic}
\end{figure*}

\begin{figure*}
\centering
\includegraphics[trim =0cm 0cm 0cm 0cm, clip,scale=.7]{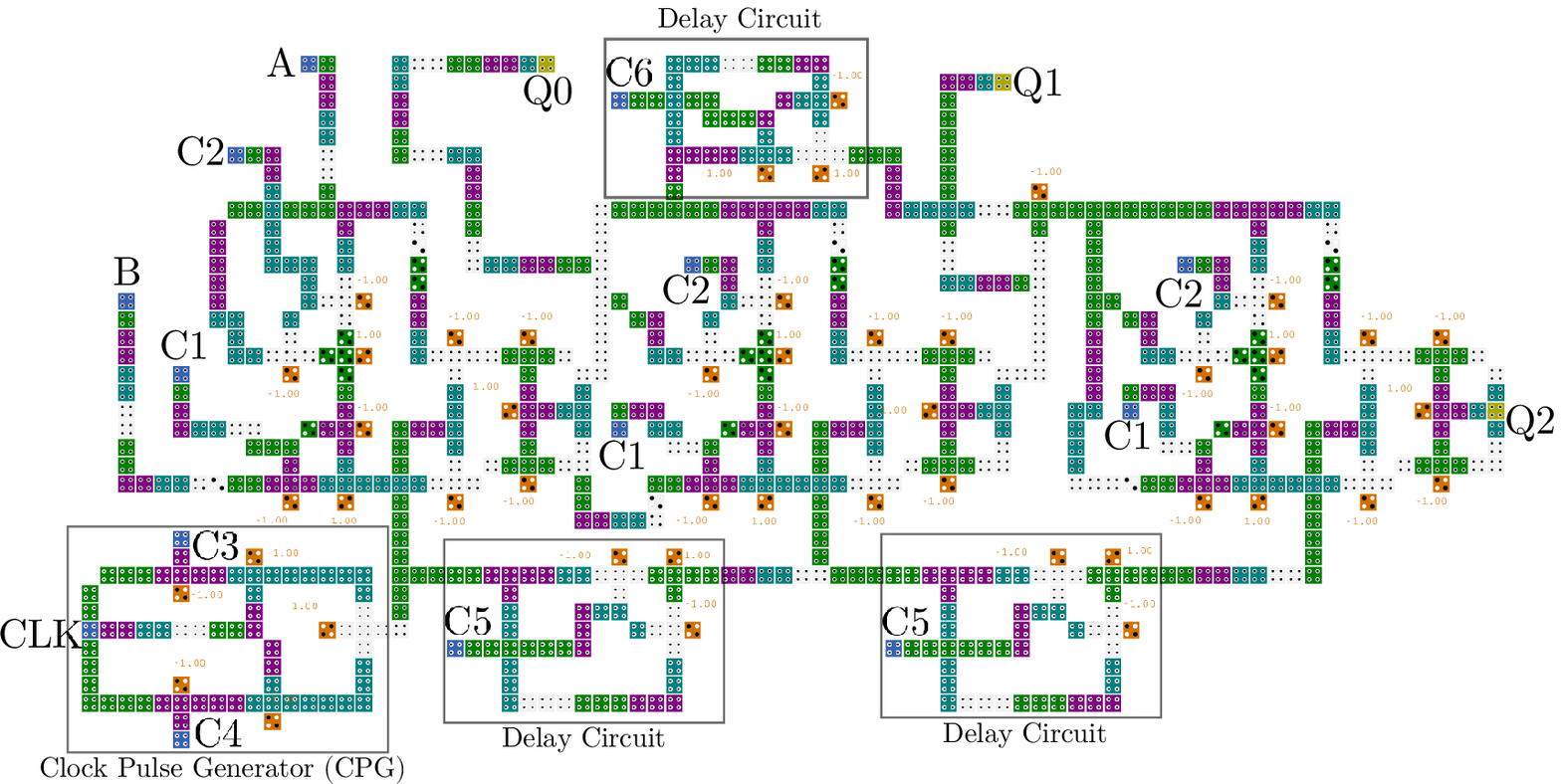}
\caption{The proposed QCA edge configurable 3-bit counter/shift register}
\label{3-bitDET}
\end{figure*}

\begin{figure*}
\centering
\includegraphics[trim =0cm 0cm 0cm 0cm, clip,scale=0.9]{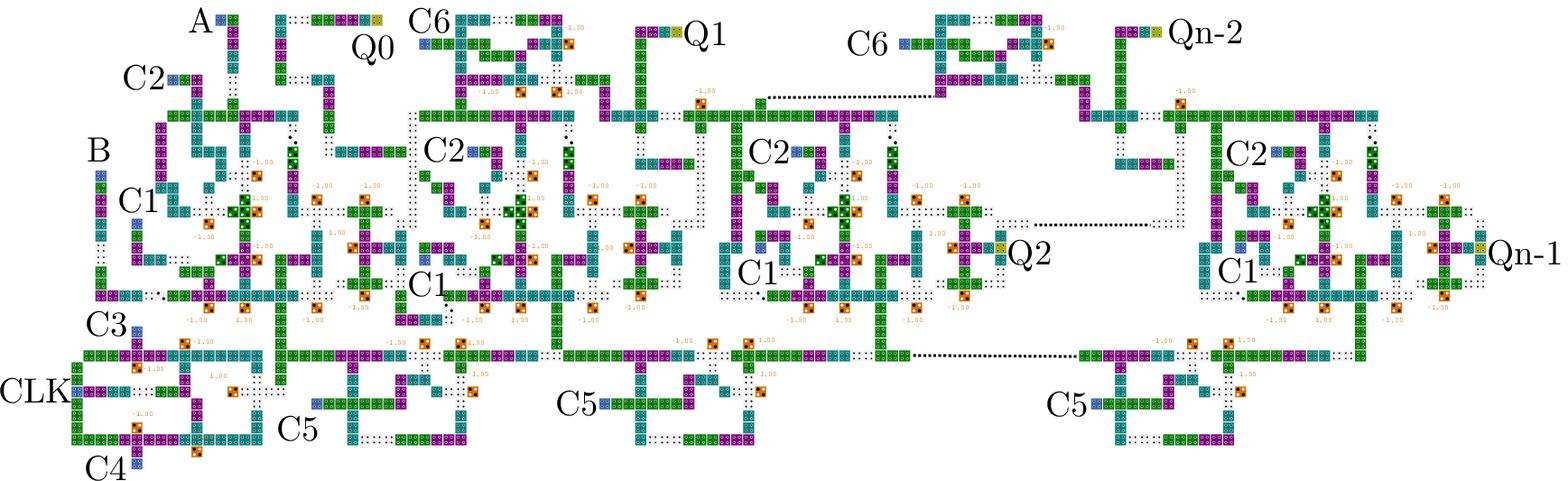}
\caption{The proposed QCA edge configurable n-bit counter/shift register}
\label{n-bitDET}
\end{figure*}

 The 3-bit design of the proposed configurable counter/ shift register is shown in Figure \ref{3-bitDET}. To synchronize the input and output of the 3-bit configurable counter/ shift register circuit, additional three delay control circuits need to be added to the ECFF module. The top delay control circuit is used to synchronize the output of the first module with rest modules whereas the bottom two delay control circuits are used to synchronize clock signal (CLK) of the proposed circuit. In the case of a 3-bit proposed edge configurable counter/ shift register, if C5=C6=0 then the circuit behaves as a shift register whereas if C5=C6=1 then the circuit will behave as a counter. Similarly, by cascading n-CFF modules, we can construct n-bit edge configurable counter/ shift register as shown in Figure \ref{schematic} which can be configured to (dual/rising/falling) edge shift register as well as (dual/rising/falling) edge counter. The QCA representation of the proposed n-bit edge configurable counter/ shift register is shown in Figure \ref{n-bitDET}. The performance of the proposed edge configurable counter/shift register is shown in Table \ref{CRDET}. If we want to construct a shift register then there is no extra space requirement for the proposed configurable counter/ shift register due to its configurable nature hence it is more cost effective than the conventional designs. The main benefit of the configurable n-bit counter/shift register is that the same circuit can be configured to six different forms, (Dual/Falling/Rising) counter as well as (Dual/Falling/Rising) shift register. 
\begin{table}
\centering
\caption{Performance of edge configurable counter/shift register}
\scalebox{.65}{
\begin{tabular}{|c|c|c|c|c|c|}\hline 
Design & Area  & Cells & Clock &Single & Configurable \\
       & $\mu m^{2}$ & & Cycle &Layer & \\\hline
  \multicolumn{6}{|c|}{Existing 2-bit Counter}\\\hline
 In \cite{DETJK} & 0.74 & 430 & 4 &Yes & No\\\hline
 In \cite{Sheikhfaal2015}        & 0.26 & 240 & 2 & No & No \\\hline
 In \cite{ang2015}            & 0.22 & 141 & 2.25 & Yes & No\\\hline
Proposed 2-bit Counter/shift register & 0.67 & 464 & 5.75 & Yes & Yes \\\hline
\multicolumn{6}{|c|}{Existing 3-bit Counter}\\\hline
In \cite{DETJK} & 1.02 & 677 & 6 &Yes & No\\\hline
In \cite{Sheikhfaal2015}         & 0.48 & 428 & 2 & No & No \\\hline
In \cite{ang2015} 			 & 0.36 & 328 & 2.25 & Yes & No \\\hline
In \cite{Abutaleb2017}           & 0.22 & 196 & 2 & Yes & No \\\hline
Proposed 3-bit Counter/shift register & 1.18 & 786 & 7.75 &Yes & Yes \\\hline
\end{tabular}}
\label{CRDET}
\end{table}
\section{Simulation setup}\label{Sim}
QCADesigner (version 2.0.3) \cite{QCAD} has been used to verify the functional correctness of all the proposed QCA designs using both coherence
vector and bistable approximation simulation engines using all the default parameters. 
\section{Conclusion}\label{Con}
This paper attempts for the very first time, to design a level triggered configurable flip-flop (CFF) which can be used as a D, T and JK flip-flop as per the requirement. A 1-bit RAM is designed considering CFF as a basic element. Moreover, an edge configurable flip-flop (ECFF) is designed with the help of a clock pulse generator (CPG). The CPG can produce three different (rising/falling and dual) types of pulses using the same circuit. The edge configurable flip-flop (ECFF) can be used as 9 different flip-flops controlling only single module. An efficient edge configurable n-bit counter/shift register is proposed which can be configured as an n-bit counter or n-bit shift register as per the requirement. 
\section*{Acknowledgement}
This research is supported by the Department of Electronics and Information Technology, Ministry of Communications and IT, Government of India under the Visvesvaraya PhD Scheme administered by Media Lab Asia.
\balance
\bibliographystyle{spmpsci}
\bibliography{jettaref}

\begin{thebibliography}{10}
\providecommand{\url}[1]{{#1}}
\providecommand{\urlprefix}{URL }
\expandafter\ifx\csname urlstyle\endcsname\relax
  \providecommand{\doi}[1]{DOI~\discretionary{}{}{}#1}\else
  \providecommand{\doi}{DOI~\discretionary{}{}{}\begingroup
  \urlstyle{rm}\Url}\fi

\bibitem{itrs}
International technology roadmap for semiconductors (itrs) (2015)

\bibitem{abedi01}
Abedi, D., Jaberipur, G., Sangsefidi, M.: Coplanar full adder in quantum-dot
  cellular automata via clock-zone-based crossover.
\newblock IEEE Transactions on Nanotechnology \textbf{14}(3), 497--504 (2015)

\bibitem{Abutaleb2017}
Abutaleb, M.: Robust and efficient quantum-dot cellular automata synchronous
  counters.
\newblock Microelectronics Journal \textbf{61}, 6 -- 14 (2017)

\bibitem{mag02}
Alam, M.T., DeAngelis, J., Putney, M., Hu, X.S., Porod, W., Niemier, M.,
  Bernstein, G.H.: Clocking scheme for nanomagnet qca.
\newblock In: 2007 7th IEEE Conference on Nanotechnology (IEEE NANO), pp.
  403--408 (2007).
\newblock \doi{10.1109/NANO.2007.4601219}

\bibitem{comT}
Angizi, S., Moaiyeri, M.H., Farrokhi, S., Navi, K., Bagherzadeh, N.: Designing
  quantum-dot cellular automata counters with energy consumption analysis.
\newblock Microprocessors and Microsystems \textbf{39}(7), 512 -- 520 (2015).
\newblock \doi{http://dx.doi.org/10.1016/j.micpro.2015.07.011}

\bibitem{ang2015}
Angizi, S., Moaiyeri, M.H., Farrokhi, S., Navi, K., Bagherzadeh, N.: Designing
  quantum-dot cellular automata counters with energy consumption analysis.
\newblock Microprocessors and Microsystems \textbf{39}(7), 512 -- 520 (2015)

\bibitem{Angizi2015}
Angizi, S., Sarmadi, S., Sayedsalehi, S., Navi, K.: Design and evaluation of
  new majority gate-based \{RAM\} cell in quantum-dot cellular automata.
\newblock Microelectronics Journal \textbf{46}(1), 43 -- 51 (2015)

\bibitem{FF5}
Angizi, S., Sayedsalehi, S., Roohi, A., Bagherzadeh, N., Navi, K.: Design and
  verification of new n-bit quantum-dot synchronous counters using majority
  function-based jk flip-flops.
\newblock Journal of Circuits, Systems and Computers \textbf{24}(10), 1550,153
  (2015).
\newblock \doi{10.1142/S0218126615501534}

\bibitem{mole02}
Awais, M., Vacca, M., Graziano, M., Roch, M.R., Masera, G.: Quantum dot
  cellular automata check node implementation for ldpc decoders.
\newblock IEEE Transactions on Nanotechnology \textbf{12}(3), 368--377 (2013).
\newblock \doi{10.1109/TNANO.2013.2251422}

\bibitem{com2T}
Bhavani, K.S., Alinvinisha, V.: Utilization of qca based t flip flop to design
  counters.
\newblock In: Innovations in Information, Embedded and Communication Systems
  (ICIIECS), 2015 International Conference on, pp. 1--6 (2015).
\newblock \doi{10.1109/ICIIECS.2015.7193059}

\bibitem{book1}
Bobda, C.: Introduction to Reconfigurable Computing.
\newblock Springer (2007)

\bibitem{bondalapati01}
Bondalapati, K., Prasanna, V.K.: Reconfigurable computing systems.
\newblock Proceedings of the IEEE \textbf{90}(7), 1201--1217 (2002).
\newblock \doi{10.1109/JPROC.2002.801446}

\bibitem{USE}
Campos, C.A.T., Marciano, A.L., Neto, O.P.V., Torres, F.S.: Use: A universal,
  scalable, and efficient clocking scheme for qca.
\newblock IEEE Transactions on Computer-Aided Design of Integrated Circuits and
  Systems \textbf{35}(3), 513--517 (2016).
\newblock \doi{10.1109/TCAD.2015.2471996}

\bibitem{WC1}
Chaudhary, A., Chen, D.Z., Hu, X.S., Niemier, M.T., Ravichandran, R., Whitton,
  K.: Fabricatable interconnect and molecular qca circuits.
\newblock Trans. Comp.-Aided Des. Integ. Cir. Sys. \textbf{26}(11), 1978--1991
  (2007).
\newblock \doi{10.1109/TCAD.2007.906467}

\bibitem{thesis01}
Chilakam, M.: A novel reconfiguration scheme in quantum-dot cellular automata
  for energy efficient nanocomputing.
\newblock Master's thesis, University of Massachusetts Amhers (2013)

\bibitem{WC100}
Devadoss, R., Paul, K., Balakrishnan, M.: Coplanar qca crossovers.
\newblock Electronics Letters \textbf{45}(24), 1234--1235 (2009).
\newblock \doi{10.1049/el.2009.2819}

\bibitem{usp}
DiLabio, G., Wolkow, R., Pitters, J., Piva, P.: Atomistic quantum dot (2014).
\newblock \urlprefix\url{http://www.google.co.in/patents/US8816479}.
\newblock US Patent 8,816,479

\bibitem{FF4}
Hashemi, S., Navi, K.: New robust \{QCA\} d flip flop and memory structures.
\newblock Microelectronics Journal \textbf{43}(12), 929 -- 940 (2012).
\newblock \doi{http://dx.doi.org/10.1016/j.mejo.2012.10.007}

\bibitem{Karkaj2017}
Karkaj, E.T., Heikalabad, S.R.: A testable parity conservative gate in
  quantum-dot cellular automata.
\newblock Superlattices and Microstructures \textbf{101}, 625 -- 632 (2017).
\newblock \doi{https://doi.org/10.1016/j.spmi.2016.08.054}

\bibitem{clock10}
Lent, C.S., Liu, M., Lu, Y.: Bennett clocking of quantum-dot cellular automata
  and the limits to binary logic scaling.
\newblock Nanotechnology \textbf{17}(16), 4240 (2006)

\bibitem{lent01}
Lent, C.S., Tougaw, P.D., Porod, W., Bernstein, G.H.: Quantum cellular
  automata.
\newblock Nanotechnology \textbf{4}(1), 49 (1993)

\bibitem{com00}
Lim, L.A., Ghazali, A., Yan, S.C.T., Fat, C.C.: Sequential circuit design using
  quantum-dot cellular automata (qca).
\newblock In: Circuits and Systems (ICCAS), 2012 IEEE International Conference
  on, pp. 162--167 (2012)

\bibitem{metel01}
Liu, M., Lent, C.S.: High-speed metallic quantum-dot cellular automata.
\newblock In: Nanotechnology, 2003. IEEE-NANO 2003. 2003 Third IEEE Conference
  on, vol.~2, pp. 465--468 vol. 2 (2003).
\newblock \doi{10.1109/NANO.2003.1230946}

\bibitem{semi01}
Mitic, M., Cassidy, M.C., Petersson, K.D., Starrett, R.P., Gauja, E., Brenner,
  R., Clark, R.G., Dzurak, A.S., Yang, C., Jamieson, D.N.: Demonstration of a
  silicon-based quantum cellular automata cell.
\newblock Applied Physics Letters \textbf{89}(1), 013503 (2006).
\newblock \doi{http://dx.doi.org/10.1063/1.2219128}

\bibitem{momen02}
Momenzadeh, M., Huang, J., Tahoori, M.B., Lombardi, F.: Characterization, test,
  and logic synthesis of and-or-inverter (aoi) gate design for qca
  implementation.
\newblock IEEE Transactions on Computer-Aided Design of Integrated Circuits and
  Systems \textbf{24}(12), 1881--1893 (2005)

\bibitem{defect}
Momenzadeh, M., Ottavi, M., Lombardi, F.: Modeling qca defects at
  molecular-level in combinational circuits.
\newblock In: 20th IEEE International Symposium on Defect and Fault Tolerance
  in VLSI Systems (DFT'05), pp. 208--216 (2005)

\bibitem{motameni01}
Motameni, H., Montazeri, B.: Reconfigurable logic based on quantum-dot cellular
  automata.
\newblock Journal of Applied Sciences Research \textbf{7}, 1817--1823 (2011)

\bibitem{Navi01}
Navi, K., Mohammadi, H., Angizi, S.: A novel quantum-dot cellular automata
  reconfigurable majority gate with 5 and 7 inputs support.
\newblock Journal of Computational and Theoretical Nanoscience \textbf{12}(3),
  399--406 (2015)

\bibitem{Navi02}
Navi, K., Roohi, A., Sayedsalehi, S.: Designing reconfigurable quantum-dot
  cellular automata logic circuits.
\newblock Journal of Computational and Theoretical Nanoscience \textbf{10}(5),
  1137--1146 (2013)

\bibitem{mole01}
Pulimeno, A., Graziano, M., Sanginario, A., Cauda, V., Demarchi, D., Piccinini,
  G.: Bis-ferrocene molecular qca wire: Ab initio simulations of fabrication
  driven fault tolerance.
\newblock IEEE Transactions on Nanotechnology \textbf{12}(4), 498--507 (2013).
\newblock \doi{10.1109/TNANO.2013.2261824}

\bibitem{Roohi03}
Roohi, A., Sayedsalehi, S., Khademolhosseini, H., Navi, K.: Design and
  evaluation of a reconfigurable fault tolerant quantum-dot cellular automata
  gate.
\newblock Journal of Computational and Theoretical Nanoscience \textbf{10}(2),
  380--388 (2013)

\bibitem{com2}
Sabbaghi-Nadooshan, R., Kianpour, M.: A novel qca implementation of mux-based
  universal shift register.
\newblock Journal of Computational Electronics \textbf{13}(1), 198--210 (2014).
\newblock \doi{10.1007/s10825-013-0500-9}

\bibitem{FF1}
Shamsabadi, A.S., Ghahfarokhi, B.S., Zamanifar, K., Movahedinia, N.: Applying
  inherent capabilities of quantum-dot cellular automata to design: D flip-flop
  case study.
\newblock Journal of Systems Architecture \textbf{55}(3), 180 -- 187 (2009).
\newblock \doi{http://dx.doi.org/10.1016/j.sysarc.2008.11.001}.
\newblock Challenges in self-adaptive computing (Selected papers from the
  Aether-Morpheus 2007 workshop)

\bibitem{Sheikhfaal2015}
Sheikhfaal, S., Navi, K., Angizi, S., Navin, A.H.: Designing high speed
  sequential circuits by quantum-dot cellular automata: Memory cell and counter
  study.
\newblock Quantum Matter \textbf{4}(2), 190--197 (2015)

\bibitem{shin}
Shin, S.H., Jeon, J.C., Yoo, K.Y.: Wire-crossing technique on quantum-dot
  cellular automata.
\newblock 2nd International Conference on Next Generation Computer and
  Information Technology \textbf{27}, 52-- 57 (2013)

\bibitem{com1T}
Torabi, M.: A new architecture for t flip flop using quantum-dot cellular
  automata.
\newblock In: Quality Electronic Design (ASQED), 2011 3rd Asia Symposium on,
  pp. 296--300 (2011).
\newblock \doi{10.1109/ASQED.2011.6111764}

\bibitem{mag01}
Vacca, M., Graziano, M., Zamboni, M.: Majority voter full characterization for
  nanomagnet logic circuits.
\newblock IEEE Transactions on Nanotechnology \textbf{11}(5), 940--947 (2012).
\newblock \doi{10.1109/TNANO.2012.2207965}

\bibitem{FF2}
Vankamamidi, V., Ottavi, M., Lombardi, F.: A serial memory by quantum-dot
  cellular automata (qca).
\newblock IEEE Transactions on Computers \textbf{57}(5), 606--618 (2008).
\newblock \doi{10.1109/TC.2007.70831}

\bibitem{comJK}
Venkataramani, P., Srivastava, S., Bhanja, S.: Sequential circuit design in
  quantum-dot cellular automata.
\newblock In: 2008 8th IEEE Conference on Nanotechnology, pp. 534--537 (2008).
\newblock \doi{10.1109/NANO.2008.159}

\bibitem{com0}
Vetteth, A., Walus, K., Dimitrov, V., Jullien, G.: Quantum-dot cellular
  automata of flip-flops.
\newblock In: ATIPS Laboratory 2500 University Drive, NW, Calgary, Alberta,
  Canada T2N 1N4 (2003)

\bibitem{QCAD}
Walus, K., Dysart, T.J., Jullien, G.A., Budiman, R.A.: Qcadesigner: a rapid
  design and simulation tool for quantum-dot cellular automata.
\newblock IEEE Transactions on Nanotechnology \textbf{3}(1), 26--31 (2004).
\newblock \doi{10.1109/TNANO.2003.820815}

\bibitem{clock11}
Wang, Y., Lieberman, M.: Thermodynamic behavior of molecular-scale quantum-dot
  cellular automata (qca) wires and logic devices.
\newblock IEEE Transactions on Nanotechnology \textbf{3}(3), 368--376 (2004)

\bibitem{DETJK}
Wu, C.B., Xie, G.J., Xiang, Y.L., Lv, H.J.: Design and simulation of dual-edge
  triggered sequential circuits in quantum-dot cellular automata.
\newblock Journal of Computational and Theoretical Nanoscience \textbf{11}(7),
  1620--1626 (2014).
\newblock \doi{doi:10.1166/jctn.2014.3541}

\bibitem{Xiao2012}
rong Xiao, L., xiong Chen, X., yan Ying, S.: Design of dual-edge triggered
  flip-flops based on quantum-dot cellular automata.
\newblock Journal of Zhejiang University SCIENCE C \textbf{13}(5), 385--392
  (2012).
\newblock \doi{10.1631/jzus.C1100287}

\bibitem{xiao2013}
Xiao, L.R., Xu, X., Ying, S.Y.: Dual-edge triggered t flip-flop structure using
  quantum-dot cellular automata.
\newblock In: Nanotechnology and Precision Engineering, \emph{Advanced
  Materials Research}, vol. 662, pp. 562--567. Trans Tech Publications (2013).
\newblock \doi{10.4028/www.scientific.net/AMR.662.562}

\bibitem{FF3}
Yang, X., Cai, L., Zhao, X.: Low power dual-edge triggered flip-flop structure
  in quantum dot cellular automata.
\newblock Electronics Letters \textbf{46}(12), 825--826 (2010).
\newblock \doi{10.1049/el.2010.1090}

\bibitem{FF6}
Yang, X., Cai, L., Zhao, X., Zhang, N.: Design and simulation of sequential
  circuits in quantum-dot cellular automata: Falling edge-triggered flip-flop
  and counter study.
\newblock Microelectronics Journal \textbf{41}(1), 56 -- 63 (2010).
\newblock \doi{http://dx.doi.org/10.1016/j.mejo.2009.12.008}

\end{thebibliography}
\end{document}